\documentclass{ieeeaccess}
\usepackage{cite}
\usepackage{amsmath,amssymb,amsfonts}
\usepackage{graphicx}
\usepackage{textcomp}
\usepackage{calc}
\usepackage{amsthm}%
\usepackage{mathrsfs}%
\usepackage{manyfoot}%
\usepackage{algorithm}%
\usepackage{algpseudocode}%
\usepackage{listings}%
\usepackage{booktabs}
\usepackage{dcolumn}
\usepackage{bm}
\usepackage{float}
\usepackage{diagbox}
\usepackage{multirow}
\usepackage{physics}
\usepackage{hyperref}
\usepackage{xcolor,colortbl}

\usepackage[normalem]{ulem}

\definecolor{lightgray}{gray}{0.9}
\definecolor{LightCyan}{rgb}{0.88,1,1}

\usepackage{easybmat}
 
\usepackage{bbold}
\usepackage{subcaption}
\def\BibTeX{{\rm B\kern-.05em{\sc i\kern-.025em b}\kern-.08em
    T\kern-.1667em\lower.7ex\hbox{E}\kern-.125emX}}
\begin{document}
\history{Date of publication xxxx 00, 0000, date of current version xxxx 00, 0000.}
\doi{10.1109/TQE.2025.DOI}

\title{Variational Quantum Eigensolver: A Comparative Analysis of Classical and Quantum Optimizer Methods}
\author{\uppercase{Duc-Truyen Le}\authorrefmark{1},
\uppercase{Vu-Linh Nguyen}\authorrefmark{2}, \uppercase{Ha C. Nguyen}\authorrefmark{3,4,5}, \uppercase{Hung Q. Nguyen}\authorrefmark{6}, \uppercase{Van-Duy Nguyen}\authorrefmark{7,8}}
\address[1]{Department of Physics, National Tsing Hua University, Hsinchu,
Taiwan 300044, R.O.C.}
\address[2]{Department of Theoretical Physics, University of Science, Vietnam
National University, Ho Chi Minh City 70000, Vietnam}
\address[3]{Département de Physique de l’École normale supérieure, Université PSL, Paris 75005, France}
\address[4]{Telecom Paris, Institut Polytechnique de Paris, 19 Place Marguerite
Perey, Palaiseau 91120, France.}
\address[5]{Institute for Quantum Computing Analytics (PGI-12), Forschungszentrum Jülich, 52425 Jülich, Germany}

\address[6]{Nano and Energy Center, University of Science, Vietnam National
University, Hanoi, Vietnam}
\address[7]{Phenikaa School of Computing, Phenikaa University, Hanoi
12116, Vietnam}
\address[8]{Phenikaa Institute for Advanced Study, Phenikaa University, Hanoi
12116, Vietnam}

\markboth
{Author \headeretal: Preparation of Papers for IEEE Transactions on Quantum Engineering}
{Author \headeretal: Preparation of Papers for IEEE Transactions on Quantum Engineering}

\corresp{Corresponding author: Van-Duy Nguyen (email: duy.nguyenvan@phenikaa-uni.edu.vn)}

\begin{abstract}
In this study, we investigated the Variational Quantum Eigensolver (VQE) application for the Ising model as a testbed model, in which we thoroughly delved into several optimizers, both classical and quantum, and analyzed the extent to which each of these methods would offer a benefit. We then investigated a new combinatorial optimization scheme, termed QN-SPSA+PSR, in which the Fubini-Study metric is approximated within the Quantum Natural Gradient (QN) framework, with its inner gradient estimated by the Simultaneous Perturbation Stochastic Approximation (SPSA), while the outer gradient of the cost function is evaluated exactly by the Parameter-Shift Rule (PSR). The QN-SPSA+PSR method integrates the QN-SPSA computational efficiency with the precise gradient computation of the PSR, improving the stability of QN-SPSA-based and convergence speed per parameter update while maintaining low computational consumption. Our results provide a potential performance improvement in the VQAs' optimization subroutine, even in Quantum Machine Learning's optimization section, and enhance viable paths toward efficient quantum simulations on Noisy Intermediate-Scale Quantum Computing (NISQ) devices. Additionally, we also conducted a detailed study of quantum circuit ansatz structures in order to find the one that would work best with the Ising model and NISQ, in which we utilized the properties of the investigated model.
\end{abstract}


\titlepgskip=-15pt

\maketitle


\section{\label{sec:level1}Introduction}
Since the early studies of quantum computers, their very quantum nature is promising for a bright new age of computation, built on three crucial features of quantum theory: quantum probabilistic behavior, superposition, and entanglement. The distinct properties of quantum computers (QC) from classical computers make them distinct in application, not only be an upgraded computational speed version of traditional computers. Currently, when the quantum hardware is rather limited on operational stability and noise resilience, despite the ambiguity of the realization of such advantages, Variational Quantum Algorithms (VQAs) \cite{cerezo2021variational, PhysRevA.92.042303} are thought to be the best at outperforming conventional computers, which are able to implement well on near-term quantum devices known as the so-called Noisy Intermediate-Scale Quantum (NISQ) computers \cite{Preskill_2018, Brooks, Zhong_2020, Arute:2019zxq, Wu_2021}.

Quantum-computing hardware has progressed rapidly, with major technology companies such as IBM, Google, and D-Wave competing to develop quantum computers and achieve quantum supremacy. The first experimental demonstration of quantum supremacy was reported by the Google AI Quantum team \cite{arute2019quantum}. However, this milestone is still far from realizing the full potential of quantum computing. Current hardware faces critical challenges, including limited qubit coherence times, restricted qubit connectivity, quantum gate implementations, and a small number of available qubits. These limitations prevent the successful execution of deep quantum circuits and the production of significant results on today’s noisy devices, which continue to be challenging to hardware scientists in the near future, posing obstacles that computer scientists would need to weigh up against trainability, precision, and efficiency. Nevertheless, NISQ devices remain exploitable for practical applications; for a comprehensive review, see Ref.~\cite{bharti2021noisy}. Standing out among quantum algorithms envisioned to beat classical computers, which are mostly designed for the fault-tolerant quantum computer, VQAs are regarded as an appropriate candidate compatible with the current defective quantum device to address these constraints \cite{PhysRevA.92.042303}. The first two prominent applications of VQAs coming up are the Variational Quantum Eigensolver (VQE) \cite{peruzzo2014variational} and the Quantum Approximate Optimization Algorithm (QAOA) \cite{farhi2014quantumapproximateoptimizationalgorithm}. In more recent developments, Quantum Machine Learning (QML) emerges as a promising application of VQAs \cite{Schuld_2014}. These inherit the core scheme of a VQA, which is to utilize the hybrid routine, leverage the versatility and computational power of classical computers to handle the computation processing, and use quantum devices to execute the quantum circuits. In this work, we examine the standard procedures of VQE and investigate the properties of different optimization methods implemented in the VQE's workflow.

The structure of this paper is as follows: in Section 2, we provide an overview of the Variational Quantum Eigensolver (VQE) procedures, a brief introduction to the Transverse Ising Model, and discuss its symmetries, which will be utilized to inform the ansatz structure. Subsequently, Section 3 delves into optimizers, encompassing both classical and quantum approaches, where we will evaluate new quantum-informed optimization methods. The numerical study and experiment are conducted in Section 4, and then we make the final conclusion in Section 5.
\section{Variational Quantum Eigensolver for Ising model}
\subsection{Overview}

\begin{figure*}
    \centering
    \includegraphics[width=\textwidth]{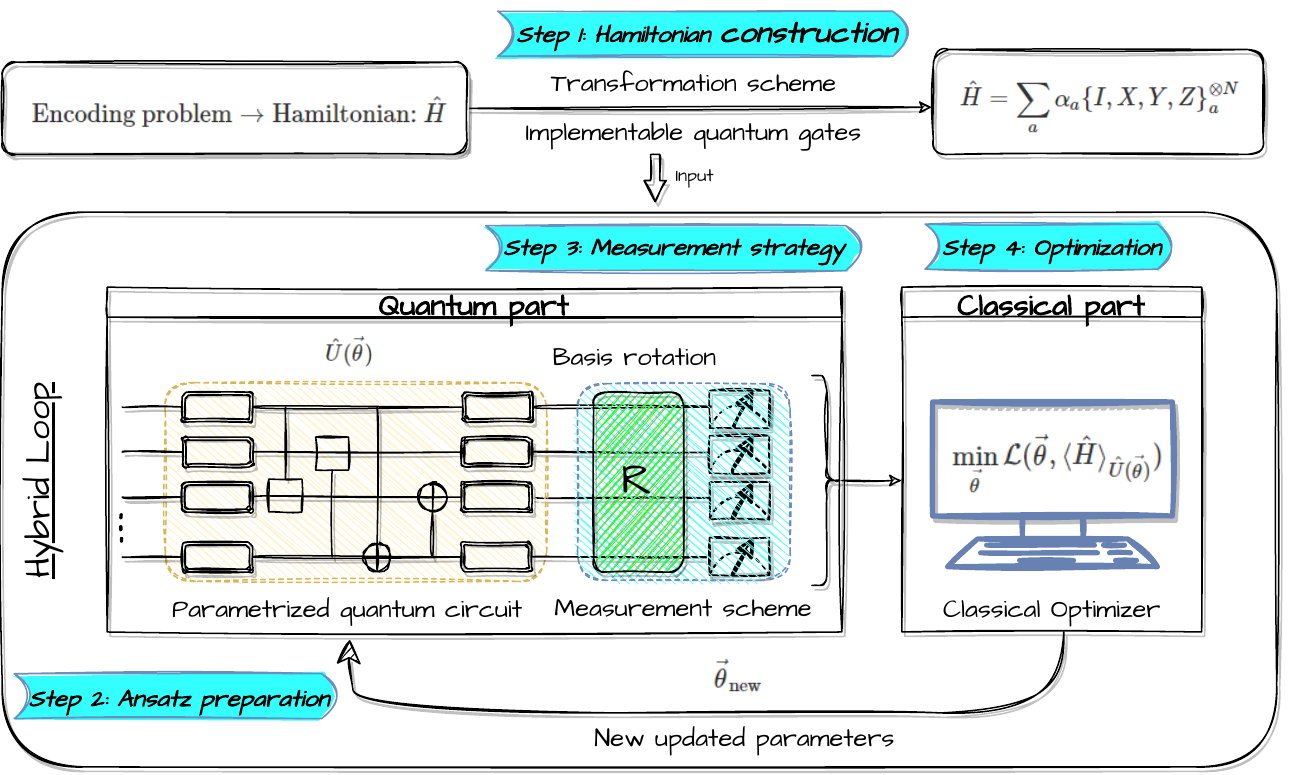}
    \caption{Variational Quantum Eigensolver (VQE) Architecture. This schematic representation illustrates the workflow of the VQE algorithm, which is a hybrid quantum-classical optimizer technique used to estimate the ground state energy of a given Hamiltonian. The process consists of four key steps: Hamiltonian Construction (Step 1): The problem is encoded into a Hamiltonian $\hat{H}$, expressed as a weighted sum of Pauli operator products acting on N qubits; Ansatz Preparation (Step 2): A parametrized quantum circuit is constructed by unitary operator $\hat{U}({\boldsymbol\theta})$, where $\boldsymbol\theta$ represents tunable parameters; Measurement Strategy (Step 3): The expectation value $\langle\hat{H}\rangle$ is obtained by applying a basis rotation followed by a specific measurement scheme; Optimization (Step 4): The measured expectation value is fed into a classical optimizer, which minimizes energy $\langle\hat{H}\rangle_{\hat{U}({\boldsymbol\theta})}$ by updating the parameters $\boldsymbol\theta$, this step is the crucial point that we will dig into details within this study. The Hybrid Loop, where quantum computations (state preparation and measurement) are performed on a quantum processor, while classical optimization is handled by a classical computer. The updated parameters $\boldsymbol\theta_{\text{new}}$ are fed back into the quantum circuit iteratively.}
    \label{VQEdiagram}
\end{figure*}

VQE was initially proposed by Peruzzo et al. \cite{peruzzo2014variational,mcclean2016theory} as an efficient alternative way to compute the ground state energy, or extended to estimate the higher-energy levels \cite{PhysRevA.95.042308,RevModPhys.40.153, Higgott_2019, PhysRevResearch.1.033062, PhysRevApplied.11.044092}, of  quantum chemistry systems \cite{doi:10.1021/acs.jpclett.3c03159, Cao_2022} or many-body systems, i.e. H${}_2$ \cite{PhysRevX.6.031007,PhysRevX.8.011021}, LiH and BeH${}_2$ \cite{Kandala_2017} molecules. Compared to Quantum Phase Estimation (QPE) \cite{NielQuant} which requires an impractically large number of quantum gates, VQE provides a much more hardware-friendly approach. Additionally, applications of VQE in drug discovery \cite{8585034, Blunt_2022} and materials \cite{Lordi_2021, Cao_2019, Bauer_2020} have attracted significant interest. Consequently, VQE serves as a flexible hybrid framework with current potential quantum resources that can outpace conventional computers \cite{Boixo_2018, mccaskey2019}. Following the success of running on the photonic quantum processor combined with the traditional devices, many research works favor the development of various types of VQA reviewed in Ref. \cite{tilly2021variational, bharti2021noisy}. In essence, what makes VQE (generally VQAs) compelling is the trial state analysis step to choose an ansatz adaptive to particular quantum device architecture and/or problems. Subsequently, the suitable ansatz is executed on the NISQ device, the remainders are handled by the classical computer through a hybrid loop. Error mitigation techniques are applied to improve the results \cite{PhysRevLett.119.180509, RevModPhys.95.045005,bharti2021noisy}. Instead of diagonalizing a matrix representation of Hamiltonian $\hat{H}$ to find out eigenvalues that exponentially scale up the matrix size and calculation overhead in traditional computing as the problem grows up larger, VQE estimates the ground state energy $E_g$, basically based on the variational principle
\begin{equation}
    E_g \leq E\left[\Psi({\boldsymbol\theta})\right] = \frac{\langle\Psi({\boldsymbol\theta})|\hat{H}|\Psi({\boldsymbol\theta})\rangle}{\langle\Psi({\boldsymbol\theta}) |\Psi({\boldsymbol\theta})\rangle} = \langle\hat{H}\rangle_{\hat{U}({\boldsymbol\theta})} .
\end{equation}
The arbitrary quantum state $|\Psi({\boldsymbol\theta})\rangle \equiv \hat{U}({\boldsymbol\theta})|\Psi_0\rangle $ is a trial solution, the so-called \textit{ansatz} parametrized by a unitary operator $\hat{U}({\boldsymbol\theta})$ \cite{cerezo2021variational, tilly2021variational,peruzzo2014variational, mcclean2016theory}, so that when the parameter ${\boldsymbol\theta}$ varies, the ansatz  $|\Psi({\boldsymbol\theta})\rangle$ is readily capable of spanning on the relevant quantum space where the ground state is located. By modifying ${\boldsymbol\theta}$ value systematically, the VQE task is to navigate the spanned space to extract the ground state minimizing the energy function $ E\left[\Psi({\boldsymbol\theta})\right]$ or, in general VQA, the energy function described by $\mathcal{L}\left({\boldsymbol\theta},\langle\hat{H}\rangle_{\hat{U}({\boldsymbol\theta})}\right)$ \cite{cerezo2021variational, tilly2021variational}, and the objective is
\begin{eqnarray}
\min_{{\boldsymbol\theta}}\mathcal{L}\left({\boldsymbol\theta},\langle\hat{H}\rangle_{\hat{U}({\boldsymbol\theta})}\right).
\label{OptLagrangian}
\end{eqnarray}
In fact, the optimal wave function is not necessarily required to be close to the ground state solution. Since the energy is an expectation value, small state-preparation errors near an eigenstate can contribute only at second order to the energy error, due to the variational stationarity of the energy. VQEs are hereby friendly with noisy devices, which can work out meaningful results even though the states are disturbed by noise. These motivations are robust enough to let us dive deeper into VQE operation, ideally with a noise-free assumption in this paper's scope though, the noisy one will be addressed in a future study.

The structure of a VQE algorithm is illustrated schematically as Fig. \ref{VQEdiagram} within four core steps: \textit{Hamiltonian construction, Ansatz preparation, measurement strategy,} and \textit{Optimization}. 

\textit{Hamiltonian construction}: Starting from a given abstract problem, e.g., molecule ground state energy, shortest path, transshipment problems,  we formulate the system's Hamiltonian $\hat{H}$ . A mapping scheme is applied to convert $\hat{H}$ into a form compatible with a quantum computer (e.g Pauli operators). Leveraged from prior knowledge of the physical system, particular symmetries are drawn on to simplify the Hamiltonian, which in turn help guide the development of other VQE subroutines.

\textit{Ansatz preparation}: To select a good ansatz, an essential condition that must be satisfied is that the space spanned by the ansatz contains the desired state that extremizes the objective function, more details can be found in  Ref.~\cite{fedorov2021vqemethodshortsurvey}. An obvious one is a generic ansatz spread over all Hilbert space, a quantum circuit for that kind of ansatz is usually generated by multi-control-$U_3({\boldsymbol\theta})$ gates so that all problems could be ultimately settled without ansatz concern. However, that is not the end of the story, the computation cost of such a multi-control-$U_3({\boldsymbol\theta})$ gate is extremely high and even unfeasible when deployed on the NISQ device. As a consequence, an initial crucial step in VQE is to find an ansatz that is expressive enough to capture the important features of the wavefunction but also efficient enough to be optimized using optimization techniques. In this work, we employ the approach that excavates up symmetries and physical properties of the physical system to guide the design of the ansatz respecting those symmetries . This substantially reduces the number of training parameters and circuit depth needed in a theoretically speculative way, thereby improving the efficiency of the optimization. There are also other methods that use a hierarchical ansatz, where the circuit is divided into layers and each layer is optimized independently, and also use machine learning to automatically generate the quantum circuit \cite{Grant_2018,doi:10.1126/science.aag2302}. 

  \textit{Measurement:}
  To estimate the expectation value of the Hamiltonian, a trial state prepared by the quantum computer needs to be read out. While in the  computational basis measurement, each qubit in the output state is measured in the $Z$ basis, the measurements in other basis can be done with an appropriate local unitary rotations. In standard VQE applications with closed quantum systems, the target Hamiltonian representing physical observable is Hermitian, therefore can be decomposed into a weighted sum of Pauli strings.
    \begin{eqnarray}
        \hat{H} & = \sum_i c_i\otimes_{j=1}^N \sigma^i_j, \\
        \sigma^i_j & \in \left\{I, X, Y, Z\right\}, c_i \in \mathbb{R}. \notag
    \end{eqnarray}
A group of qubit-wise commuting Pauli strings  $P^i = \otimes_{j=1}^N \sigma^i_j$ where at each qubit-site corresponding operators are either identical $\sigma^m_j \equiv \sigma^k_j$ or identity can be measured in a single product-basis measurement setting. The grouping methods reduce measurement overhead, problem symmetries can be exploited further. A review of these and other efficient measurement methods is provided in \cite{PhysRevX.10.031064}. These active research areas are crucial to NISQ devices when a minimum measurement shots is required. Furthermore, to achieve an energy accuracy $\epsilon$, standard VQE typically requires $\mathcal{O}(1/\epsilon^2)$ circuit samples, but uses shallow circuits with depth $\mathcal{O}(1)$. In contrast, QPE can reach the same accuracy with only $\mathcal{O}(1)$ measurement repetitions, but requires much deeper circuits with depth $\mathcal{O}(1/\epsilon)$. To interpolate between these two regimes, the generalized $\alpha$-VQE framework has been proposed. This approach reduces the sampling cost of VQE to $\mathcal{O}(1/\epsilon^{2(1-\alpha)})$ while requiring circuit depth $\mathcal{O}(1/\epsilon^\alpha)$, where $\alpha \in [0,1]$ \cite{Wang_2019}. Thus, $\alpha$-VQE provides a tunable balance between the shallow-depth, sampling-intensive nature of VQE and the deep-circuit, sample-efficient nature of QPE.
    
\textit{Optimization}: The last piece in the VQE workflow is to vary the parameter ${\boldsymbol \theta}$ to find the global minimum (although in many cases, a local deep minimum is sufficient) of the function of interest described by Eq.~(\ref{OptLagrangian}). To ensure practical implementation several criteria should be considered, including, low-cost function evaluation, fast convergence with a given quantum of resources, and robustness to noise. Different ways of utilizing objective function information classify optimization approaches into two main categories: gradient-based and derivative-free. On the one hand, the gradient-based approach employs the gradient information in mostly first order, and second order to travel along the steepest direction of the cost function landscape. Conversely, the derivative-free rather directly uses responses of the objective function at different points in the parameter space to iteratively improve the parameter values. Because these utilize the classical algorithm subroutine and  consume a limited quantum evaluation expense compared to gradient-based, they are often considered more well-suited for NISQ. In the context of VQAs, gradient-free approaches can be used to optimize quantum circuit parameters when computing the gradient is either impractical or excessively costly, for example, when simulating the behavior of a quantum system becomes computationally expensive, or when the objective function cannot be expressed analytically. Nevertheless, the gradient-based methods are arguably more reliable and  faster  converging as the more complicated structure of the objective function is. Hence, a balanced trade-off should be carefully investigated  between these two approaches. Furthermore, the step size problem in gradient-based and applying machine learning in derivative-free are active topics of research at this stage, see \cite{bharti2021noisy} section II.D for more reviews. In the context of this paper, we investigated several optimization methods and instead labeled them again into two regimes: classical and quantum optimizers.

\subsection{\label{TIM} Transverse Ising model} 
The Transverse Ising model (TIM) was introduced first in 1963 by de Gennes when he built a model for describing the low-frequency collective modes of protons in the ferroelectric phase of ferroelectric crystals $\displaystyle \text{KH}_2\text{PO}_4$ \cite{DEGENNES1963132}. From then until now, the Transverse Ising Model has been used to represent enormous simple-to-complex systems such as brain science and information science and technology.

We consider the 1D Transverse Ising ring, describing the nearest-neighbor interactions of the spin projection along the z-axis and uniform external magnetic field along the x-axis (in principle, perpendicular to the z-axis). Defined by the two-spin exchange interaction factor $J$ and representative strength of the external field $h$ as described below

\begin{eqnarray} 
    {H_{{\text{TIM}}}} =  - J\mathop \sum \limits_{n = 0}^{N - 1} {\sigma^z _n}{\sigma^z _{(n + 1)\bmod N}} - h\mathop \sum \limits_{n = 0}^{N - 1} {\sigma^x_n}
    \label{Isingeq}.
\end{eqnarray}
 
The Hamiltonian in Eq.~(\ref{Isingeq}) possesses a $\mathbb{Z}_2$ symmetry remaining invariant under spin-flipping action. When the coupling constant is $h/J < 1$, the system exhibits a ferromagnetic phase, and the spins favorably align along the z-direction. Conversely, for $h/J > 1$, the system transitions to a disordered paramagnetic phase. The $\sigma_x$ terms take a dominant role when $h \rightarrow \infty$, leading the ground state to align predominantly in the $\ket{+}^{\otimes N} $ state. At the critical point, $h/J=1$ renders the gapless property in the thermodynamic limit.

 One of the key differences between the classical and quantum versions of the Ising model is that the expectation value of two terms in Eq.~(\ref{Isingeq}) must be measured separately due to the non-commutative feature of these elements. However, the commutation property allows us to perform only two observations to extract the Hamiltonian's expectation. The first term $\sum_{n=0}^{N-1}\sigma^z_n\sigma^z_{(n+1)\bmod N}$ can be cumulatively achieved by measuring directly on the computational basis as follows:

\begin{align}
    &\langle\sigma^z_m \sigma^z_n \rangle =    \sum_{\{q_0 , \cdots , q_{N-1}\} = \{0,1\}}  |a_{q_0 \cdots q_m \cdots q_n \cdots q_{N-1}}|^2    (-1)^{q_m+q_n}, \label{eq:5} \\
    &\langle \sigma^z_n \rangle = \sum_{\{q_0 , \cdots , q_{N-1}\} = \{0,1\}} |a_{q_0  \cdots q_n \cdots q_{N-1}}|^2    (-1)^{q_n},\label{eq:6}
\end{align}
with $\displaystyle |a_{q_0 \cdots q_m \cdots q_n \cdots q_{N-1}}|^2$ and $\displaystyle |a_{q_0  \cdots q_n \cdots q_{N-1}}|^2 $ are the probabilities of being in the computational basis state $\displaystyle \ket{q_0 \cdots q_m \cdots q_n \cdots q_{N-1}}$ and $\displaystyle \ket{q_0 \cdots q_n \cdots q_{N-1}}$ respectively. Similarly, $\displaystyle\sum_{n=0}^{N-1}\sigma^x_n$ is obtained by applying the Hadamard gate or the $\displaystyle R_y\left( -\frac{\pi}{2} \right)$ gate in advance to transform to the Z-basis, or computational basis, and then taking the measurements following Eq.~(\ref{eq:6}). 

\subsection{\label{ansatzconst}Ansatz Construction}
To construct an ansatz in terms of a parameterized quantum circuit (PQC), as mentioned above, we need the $U({\boldsymbol\theta})$ operator to be able to span all quantum Hilbert space so that we can use it to represent a broad class of target states. In practice, a general $U(\theta)$ operator is difficult to implement experimentally because of the noisy and weighty composite $C-U_3$ gate and actually does not acquire overall good operation even in an ideal simulator. Hence, the ansatz implementation for a particular problem's purpose is a serious studying process of the VQE working roadmap. In addition to the compatibility with the current quantum device, tapping into symmetries and properties of our problem could be leveraged to shrink the parameter space, which are the important criteria for adopting an ansatz.  
\subsubsection{Symmetry of the Transverse Ising model \label{symmetry}}
In this work, we consider three properties of TIM so that we take its suggesting information into account to decrease the size of the ansatz
\begin{itemize}
    \item \textit{Real representation}.
    Using the eigenstates of the $\sigma_Z$ (Pauli-z) operator as the elementary binary computational basis, in terms of which, $\sigma^z$, $\sigma^x$ are real matrices, due to that, we can represent the TIM Hamiltonian in real form, the eigenstates of the TIM Hamiltonian can thus be chosen to be real for conventional purposes. Namely, considering $|\Psi\rangle = \sum_n C_n|n\rangle$ is an eigenstate of $ H_{\text{TIM}}$ expanding in the computational basis $|n\rangle$. The real form of $ H_{\text{TIM}}$ means the Hermitian real element matrix that induces coefficients satisfying $C^*_n C_m = C_n C^*_m$ $\forall m,n \in [0,2^N-1]$, then, in generality  
    \begin{eqnarray}
    C_n = r_n e^{i(c+k_n\pi)}, \quad \text{$c$ is a constant,}
    \label{realre}
    \end{eqnarray}
    or, in other words, the angles in the complex plane of coefficients $C_n$ differ from every other by a factor $k\pi$, where $k$ and $k_n$ are integers. Then, the complex angle can be shifted to the real coefficient $C^*_n = C_n$ using the quantum global phase principle.  
    \item \textit{Local interaction}.
    The first term $\hat{H}_{\text{TIM}}^{\text{LI}}$ in the TIM Hamiltonian describes the kind of neighboring spin interaction along the z-axis
     \begin{eqnarray}
    \hat{H}_{\text{TIM}}^{\text{LI}} =  - J\mathop \sum \limits_{n = 0}^{N - 1} {\sigma^z _n}{\sigma^z _{(n + 1)\bmod N}}
    \end{eqnarray}
    The interaction of two local spins formulated by this term causes a sort of entanglement structure between every two local spins of the ground state energy. Nevertheless, in the case of the order phase, when the external magnetic field is dominant, this interaction can be broken down, each spin will be freely interacting and align in the same direction as the magnetic field. 
    \item \textit{Total spin-flip symmetry \label{spinflip}}. The most featured symmetry of the Ising model can be referred to as the classical counterpart, time-reversal symmetry (generally, $\mathbb{Z}_2$ symmetry). Under the total spin-flip transformation in the z direction $\left(\sigma^x\right)^{\otimes N}$ 
    \begin{equation}
        \left[\left(\sigma^x\right)^{\otimes N},\hat{H}_{\text{TIM}}\right] = 0,
    \end{equation}
    which implies the TIM Hamiltonian remains unchanged. This one leads to the eigenstate $|\Psi\rangle$ and $\left(\sigma^x\right)^{\otimes N}|\Psi\rangle = |\tilde{\Psi}\rangle$ has the same energy, the relation between them is thus put in two cases
    \begin{eqnarray}
    \langle\Psi|\tilde{\Psi}\rangle = \begin{cases}0 & \text{Degeneracy} \\ \pm 1 & \text{Non-degeneracy}\end{cases},
    \end{eqnarray}
where, in the degenerate case, it $g =0$ is trivial, and in the non-degenerate $g>0$ case, the expanded coefficients in the z-direction basis representation have a structure \begin{equation}
    C_n = \pm C_{2^N-1-n}.
    \label{spinflipcoeff}
\end{equation}
\end{itemize}
\subsubsection{Ansatz selection}
\label{Ansatz selection}
\begin{figure}[H]
    \centering
    \begin{picture}(0,0)
        \put(-120,2){\textbf{(a)}}
    \end{picture}
    \begin{subfigure}[b]{0.5\textwidth}
        \centering
        \includegraphics[width=\textwidth]{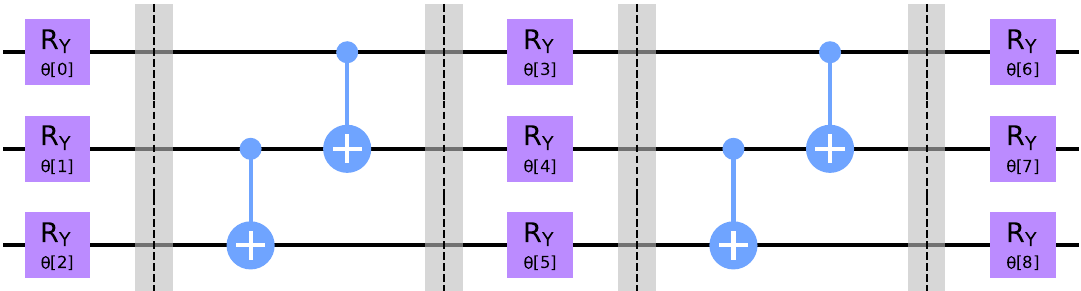}
        \label{reallin}
    \end{subfigure}
    
    \begin{picture}(0,0)
        \put(-120,2){\textbf{(b)}}
    \end{picture}
    \begin{subfigure}[b]{0.5\textwidth}
        \centering
        \includegraphics[width=\textwidth]{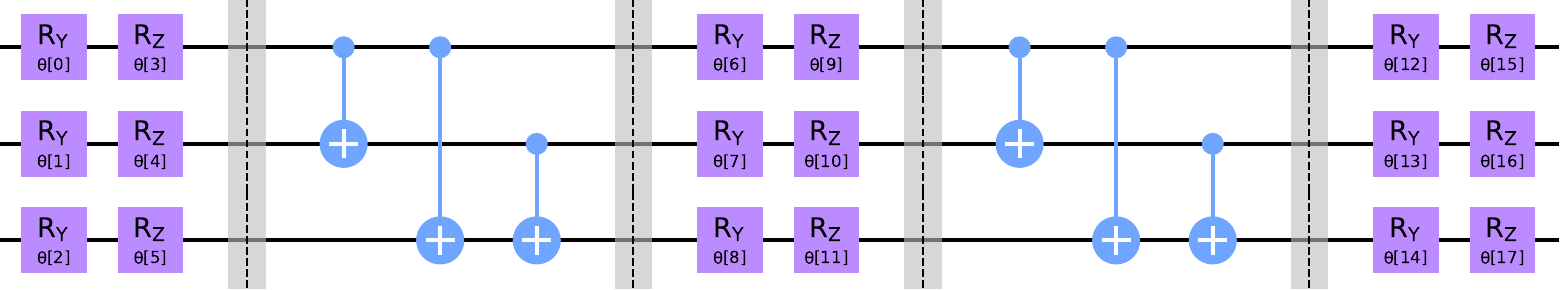}
        \label{Effsu2full}
    \end{subfigure}
    \caption{The two hardware-efficient ansatz circuits investigated in this study: a) RealAmplitudes circuit with the linear entanglement scheme and b) EfficientSU2 circuit with the full entanglement scheme. The rotation gate R${}_{\text{Y}}$ can steer state within real space. Additionally, the single-qubit SU2-group-rotation gates \{R${}_{\text{X}}$, R${}_{\text{Y}}$, R${}_{\text{Z}}$\} and CNOT gate connecting qubits are practically implemented at scale.}
    \label{realanstz}
\end{figure}

For the chosen ansatz, the conventional real coefficients of eigenstates tell us that it is enough to span in real quantum parameter space for finding the ground state energy, and the linear entanglement mapping comes from the information of the local interaction term in the Hamiltonian. Moreover, to accommodate the device constraints, in a speculative way, the available common-use RealAmplitudes ansatz is our suitable candidate for the implementation, which uses only the quantum gate $\displaystyle R_y\left( \bf \theta \right)$ for the parametrized quantum circuit and the entanglement scheme as in Fig.~\ref{realanstz}a. We also compare the linear-mapping and full-mapping entanglement schemes, as well as the RealAmplitudes and EfficientSU2 ansatz, the latter allowing quantum states to be represented in a complex-valued Hilbert space. These comparisons are discussed in Section~\ref{Simulation}.
    
    
   

The additional reason for choosing such a Hardware-efficient ansatz architecture \cite{Kandala_2017} comes from the efficient implementation of quantum optimizers to attain our full quantum algorithm purpose, which will be addressed in the next subsection \ref{quaop}. 

To efficiently navigate the full quantum Hilbert space, we need to tune $2^{N+1}-2$ degrees of freedom, $+1$ inside of the exponent indicates the complex number condition and $-2$ turns up from normalization and global phase conditions. After bringing TIM's properties, we investigated from section \ref{symmetry} to dissect. \textit{Real representation} Eq.~(\ref{realre}) helps us eliminate $+1$ within the exponent, and \textit{Total spin-flip symmetry} Eq.~(\ref{spinflipcoeff}) reduces the number of parameters we have to search the ground state by half. Finally, 
\begin{equation}
2^{N+1}-2 \rightarrow 2^{N-1}-1. 
\label{paracondi}
\end{equation}
The number of layers $L$ of the RealAmplitudes ansatz, which changes directly with the ansatz number of parameters $p$ according to $p=NL$. Following Eq.~(\ref{paracondi}), we evaluate an empirical shallow-circuit choice with the number of layers 
\begin{equation}
    L\geq \frac{2^{N-1}-1}{N},
    \label{layer estimation}
\end{equation}
. This is not a rigorous required condition because the ansatz's entanglement mapping scheme used also contributes, which is able to change the bound as well, it is still a good estimation for surface analysis, nonetheless. Yet, we can realize that, not only TIM, but also Eq.~(\ref{paracondi}) holds for all symmetrically real Hamiltonians being symmetric under $\left(\sigma^x\right)^{\otimes N}$ the operator. Therefore, there are some hidden TIM properties that we are unable to analytically incorporate to give an exact boundary value of $L$. Through conducting the experimental survey though, we are able to choose a reliable value, as we will see in Section \ref{Simulation}, it works well even for choosing $L=2$ for all qubit scaling.

\section{Optimization}
\label{opsec}
The next subroutine in the VQE work map, where the optimization of variational parameters plays a key role in hybrid quantum-classical algorithms, which requires efficient methods to achieve fast and accurate convergence.  The choice of optimization methods directly impacts the performance of the whole algorithm, influencing their ability to reach optimal solutions with minimal computational cost \cite{bharti2021noisy, tilly2021variational}. Our goal is to investigate different optimization techniques, analyze their performance to enhance computational efficiency and stability, and further evaluate them numerically in terms of convergence speed and accuracy. Henceforth, it should be noted that the convergence speed considered here does not refer to computational speed in terms of runtime, rather, it refers to the speed at which information is updated.
Effective optimizers are essential not only for VQE but also for other variational approaches for quantum chemistry, material simulations and combinatorial optimization tasks. For broader applications, in Quantum Machine Learning (QML) that plays a critical role in training quantum neural networks, kernel-based models, and generative learning algorithms\cite{Schuld_2014}. 

In this work, we focus on the Transverse Ising Model as a test bench, where we minimize the cost function defined as the expectation value of the Hamiltonian, $\langle \hat{H}_{\text{TIM}} \rangle$, using common methods of optimizer categorized into two class operations.
\subsection{\label{clasop} Classical Operation}
Classical operation means regardless of the analysis of the quantum structure of the cost function, in particular ansatz structure, we are able to find the optimal parameters minimizing the objective function. Normally, in the sense of operation, we can do the classical optimization process independently from other VQE parts, and even with  problem-specific concerns, and it appears that one can develop an optimization algorithm for general variational issues.

\textit{Constrained Optimization By Linear Approximation (COBYLA)}. One of the most powerful derivative-free methods, which has been favored by many users in recent decades, COBYLA makes use of linear interpolation of the objective function at each iteration by a unique linear polynomial function at the vertices for finding an optimal vector parameter within the trust region, then feeding the optimal point evaluated to the objective function to get the value improving the next iteration of approximation \cite{powell_1998,powell2007view,conn1997convergence,powell1994direct}. In VQE, COBYLA's linear interpolation at each step is a classical subroutine running on the CPU using only one objective value from the QPU evaluation running. And so, because of ignored derivative information, it would beneficially avoid several problems in analysis optimizations, especially Barren plateaus landscape, in return for less accuracy for more parameters ($p>9$) \cite{powell1994direct}. For a bird's-eye view, new updates to COBYLA versions like UOBYQA, NEWUOA, and BOBYQA count the objective function's curvature information to increase convergence \cite{powell2002uobyqa,powell2006newuoa,powell2009bobyqa}. 

\textit{Finite Difference (FD).} Let's dawn on one of the primary first-order derivative-based approaches, where the parameter ${\boldsymbol\theta} \in \mathbb{R}^p$ is updated at the $k$-th iteration by
\begin{equation}
    {\boldsymbol\theta}_{k+1} = {\boldsymbol\theta}_k - \eta_k \triangledown f({\boldsymbol\theta}_k),
    \label{oddupdate}
\end{equation}
which we usually name the Gradient Descent (GD) method \cite{bharti2021noisy}. Finite Difference is a numerical technique to calculate the gradient vector $\triangledown f({\boldsymbol\theta}_k)$ without the analytical function's texture, which is the basic one used in the gradient descent method. We use the well-known central difference formula,
\begin{equation}
    \triangledown f({\boldsymbol\theta}_k)_i \simeq \frac{f({\boldsymbol\theta}_k+\epsilon \vec{i})-f({\boldsymbol\theta}_k-\epsilon \vec{i})}{2\epsilon},
\end{equation}
where $\vec{i} \in \mathbb{R}^p$ is $i$-th the unit vector and $\epsilon$ is an infinitesimal change, with the error proportional to $\mathcal{O}(\epsilon^2)$. The smaller $\epsilon$ we use, the more accurate the result we get. However, due to the restricted accuracy of a classical computer, we are unable to achieve a value of $\epsilon$ that is too small, which even becomes greater when implemented on a quantum device, where at least the sample error enters the picture.

\textit{Simultaneous Perturbation Stochastic Approximation (SPSA).} To surmount some of the obstacles that arise from gradient descent operation on near-term devices, the idea of a perturbing stochastic approximation method is a good choice. SPSA generates an unbiased estimator $\tilde{f} ({\boldsymbol\theta}_k)$ of the gradient by simultaneously randomly perturbing the gradient direction of parameters \cite{spall1998overview}, we replace the ordinary gradient vector $\triangledown f({\boldsymbol\theta}_k)$ by

\begin{equation}
    \triangledown f({\boldsymbol\theta}_k) \rightarrow \triangledown \tilde{f} ({\boldsymbol\theta}_k) = \frac{f({\boldsymbol\theta}_k+s_k \vec{\Delta}_k)-f({\boldsymbol\theta}_k-s_k \vec{\Delta}_k)}{2s_k}\vec{\Delta}_k,
\end{equation}
where $\vec{\Delta}_k$ is a random perturbed vector sampled from the zero-mean distribution, usually, the Bernoulli distribution is used. We can see that all parameters are simultaneously shifted by a random amount ($\pm s_k$), our computation, therefore, only requires two objective function evaluations per iteration. Whereas standard gradient computation time is scaled along with the number of parameters, the SPSA optimizer is, however, independent, which saves a lot of time as we work on a higher parameter regime. Besides that, noise from quantum circuit executions computing the objective value can be regarded as a stochastic perturbation part absorbed into the SPSA procedure. These advantages promote SPSA and its other versions to be efficient techniques in the NISQ era.
\subsection{Quantum Operation}
\label{quaop}
As opposed to the classical ones, the quantum operations require further quantum information extracted from the structure of the problem, such as the trial wave function. This sort of property inevitably entails the details throughout the problem analysis, especially regarding Ansatz Construction.

\textit{\label{psr}Parameter-shift rules (PSR)}. Inspired by the classical shift rules in the exact derivative computation of some special functions, we are successfully able to figure out the analytical value of the objective function's derivatives using quantum devices \cite{mitarai2018quantum, Schuld_2019}. For any kind of quantum gate that has the form
\begin{equation}
    \hat{\mathcal{G}}(\theta) = e^{-i\theta \hat{G}}
\end{equation}
generated by the Hermitian operator $\hat{G}$. Suppose that we have an expectation value of $\hat{H}$, which is our objective function $f(\theta) = \langle\psi(\theta)|\hat{H}|\psi(\theta)\rangle$ parametrized by $\theta$. The ansatz wave function $|\psi(\theta)\rangle = \hat{U}(\theta)|\psi\rangle_I$ is made up of $\hat{U}(
\theta) = \hat{A}\hat{\mathcal{G}}(\theta)\hat{B}$, with $\hat{A}$, $\hat{B}$ are other arbitrary operators. As a result, the partial derivative of the $f(\theta)$ with respect to $\theta$ is thus obtained via the parameter-shift rules \cite{Schuld_2019}
\begin{equation}
    \partial_\theta f(\theta) = s \left[f(\theta+\frac{\pi}{4 s})-f(\theta-\frac{\pi}{4 s})\right].
\end{equation}
The partial derivative $\partial_\theta = \frac{\partial}{\partial \theta}$ implies we can generalize to a set of multiple parameters $\{\theta_i\}$. The evaluation of $f(\theta \pm \frac{\pi}{4 s})$ can be easily run on quantum computers, and then the exact amplitude of the vector gradient is derived. Within the aim of this paper, the given parametric ansatz is made out of Pauli rotations, and $s = \frac{1}{2}$ is selected accordingly. The state-of-the-art formalism for more generic PQCs is also able to be derived \cite{Banchi2021measuringanalytic,Wierichs2022generalparameter}.

\textit{Quantum Natural Gradient Descent (QNG).}  How to improve the convergence? The global fixed learning rate $\eta_k$ is apparently not a good choice, as one does not be good sensitive to the model information with respect to parameter changes. Initially, many attempts try to tune the learning rate $\eta_k$, such as learning rate schedulers that vary $\eta_k$ after an iteration, latterly, adaptive methods that count previous iteration values or capture the curvature of the objective function using the diagonal approximation of the Hessian (HES), where each $\eta_k$ is the inverse Hessian's diagonal element instead of being equally fixed. Those methods suggest that we can embed further information, rather than only the gradient vector, to make the optimal step size for our variational quantum algorithms. Similarly to the classical counterpart using the Fisher Information Matrix (FIM), Quantum Natural Gradient Descent invokes the quantum geometry of the wave function\cite{Stokes_2020}, which can help us get the optimal spot faster. Namely, we transform to the Riemann parameter space using the metric $g \in \mathbb{R}^{p \times p}$ of the projected Hilbert space $\mathcal{PH}$ instead of the flat Euclidean space $g=\mathbb{1}_{p\times p}$, where $p$
denotes the number of variational parameters. The update Eq. (\ref{oddupdate}) turns to
\begin{equation}
    {\boldsymbol\theta}_{k+1} = {\boldsymbol\theta}_k - \eta_k g^{+}({\boldsymbol\theta}_k) \triangledown f({\boldsymbol\theta}_k),
\end{equation}
where $g^{+}({\boldsymbol\theta}_k)$ denotes the pseudo-inverse othe the local metric $g ({\boldsymbol\theta}_k)$. The metric $g$ of the Riemann parameter space  is generally defined as
\begin{equation}
    ds^2 = g_{ij}d\theta_i d\theta_j.
\end{equation} 
The Hilbert space $\mathcal{H}$ of "bare" quantum states reduces to the $\mathcal{PH} = \mathcal{H}/U(1)$ space as we ignore the local phase $U(1)$ of quantum states \cite{https://doi.org/10.48550/arxiv.1012.1337}, the quantum distance in $\mathcal{PH}$ is then based on to calculate the Riemann metric $g$
\begin{equation}
\begin{aligned}
    ds^2 &= 1 - |\langle \psi_\theta , \psi_{\theta+d\theta}\rangle|^2  \\
    &= 1 - \left|1+\frac{1}{2}\langle \psi_\theta| \partial_i \partial_j \psi_\theta\rangle d\theta_i d\theta_j+\langle \psi_\theta| \partial_i\psi_\theta\rangle d\theta_i\right|^2\\
    & =\text{Re}[G_{ij}] d\theta_i d\theta_j.
\end{aligned}
\end{equation}
We expand up $|\psi_{\theta+d\theta}\rangle = |\psi_\theta\rangle + |\partial_i \psi_\theta\rangle d\theta_i + \frac{1}{2}|\partial_i \partial_j \psi_\theta\rangle d\theta_i d\theta_j$ up to second order in $d\theta$, where $\partial_i = \frac{\partial}{\partial \theta_i}$. The quantum geometry tensor (QGT) $G_{ij} \in \mathbb{R}^{p \times p}$ has two parts: the anti-symmetric imaginary part $\sigma_{ij} = - \sigma_{ji}$ related to the gauge field of $U(1)$ is not used in the present update rule. The contribution from the real symmetric part $g_{ij} \equiv g_{ij}({\boldsymbol\theta})$ reflects the quantum distance in $\mathcal{PH}$ space as

\begin{equation}
     g_{ij}({\boldsymbol\theta}) = \text{Re}\left[\langle \partial_i \psi_\theta| \partial_j \psi_\theta\rangle-\langle \partial_i \psi_\theta|\psi_\theta\rangle\langle \psi_\theta| \partial_j \psi_\theta\rangle\right].
     \label{qngmatrix}
\end{equation}
The Fubini-Study metric tensor $g_{ij}({\boldsymbol\theta})$ is a quantum analogue of the Fisher information matrix (QFIM). Given that it is proportional to $p^2$, the complicated computing and computational cost of $g({\boldsymbol\theta})$ is high and incompatible with the short-term quantum device. To solve this problem, an approximation strategy is required. 

\textit{Quantum Natural Gradient - Block Diagonal Approximation (QN-BDA)}. Moreover, apart from Diagonal Approximation, where we just count $p$ elements in the diagonal line, the QN-BDA employing the Parametric Family circuit is conveniently deployed entirely in the computation of the approximated metric tensor on the quantum computer \cite{Stokes_2020}. The parametric unitary operator $ \hat{U}({\boldsymbol\theta})$ acts on the initial state, $|\psi\rangle_I$ entailing $L$ layers
    \begin{equation} 
        \hat{U}({\boldsymbol\theta})= S_L P_L(\theta^{L}) \hdots S_1 P_1(\theta^1),
    \end{equation}
    where:
    \begin{itemize}
        \item $S_l$ are the static parts, which usually denote the entanglement section.
        \item $P_l(\theta^{l})$ are the parametric parts, which can be decomposed to single qubit gates $P_l(\theta^{l})=\bigotimes_{i=1}^N R_i(\theta^l_i)$.
        \item The single qubit gate $R_i(\theta^l_i) = \exp{\left[i\theta_i^l K_i\right]}$ is constructed from the Hermitian generator $K_i$ with the parameter $\theta_i^l \in \theta^l = \left\{\theta^l_1,\hdots,\theta^l_N\right\}$.
    \end{itemize}
Such a type of parametrized circuit, whose nice properties we can prospect to yield a block diagonal form of QGT running completely on the quantum processor, each block corresponding to each layered vector parameter $\theta^l \in {\boldsymbol\theta} = \theta^1 \oplus \hdots \oplus \theta^L $. We denote
\begin{eqnarray}
\hat{U}^m_{n} =  S_m P_m(\theta^{m})\hdots S_n P_n(\theta^n),
\end{eqnarray}
then $\hat{U}({\boldsymbol\theta}) \equiv \hat{U}^L_{1} = \hat{U}^L_{l+1} S_l P_l(\theta^{l})\hat{U}^{l-1}_{1}$. The partial derivative state can be written in the form
\begin{equation}
\begin{aligned}
    |\partial_i \psi_\theta\rangle & = \partial_i \hat{U}({\boldsymbol\theta}) |\psi\rangle_I = \hat{U}^L_{l+1} S_l \partial_i P_l(\theta^{l}) \hat{U}^{l-1}_{1}|\psi\rangle_I \\
   & = \hat{U}^L_{l} \left(iK_i\right) |\psi^{(l-1)}_{\theta}\rangle.
\end{aligned}
\end{equation}
Note that $\left[K_i,K_j\right] = 0$ in each layer, the unitarity of $\hat{U}^m_{n}$ and the overlap of partial derivative states tell us the block matrix element 
\begin{equation}
\begin{aligned}
    g^{(l)}_{ij}({\boldsymbol\theta}) = &\langle  \psi^{(l-1)}_\theta|  K_i K_j|\psi^{(l-1)}_\theta\rangle \\
    &-\langle \psi^{(l-1)}_\theta|K_i|\psi^{(l-1)}_\theta\rangle\langle \psi^{(l-1)}_\theta| K_j|\psi^{(l-1)}_\theta\rangle
\end{aligned}
\end{equation}
of the block-diagonal form of the Fubini-Study metric 



\[
g_{ij}(\boldsymbol\theta) =
\begin{array}{c c}
   & \begin{array}{cccc}
      \hspace{-2.5mm}  \theta^1 & \hspace{2.5mm} \theta^2 & \cdots & \hspace{2mm} \theta^L
     \end{array} \\ \hspace{-3mm}
\begin{array}{c}
   \theta^1 \\ \theta^2 \\ \vdots \\ \theta^L
\end{array}
&  \hspace{-5mm}
\left(
\begin{array}{cccc}
 g^{(1)} & 0        & \cdots & 0 \\
 0       & g^{(2)}  & \cdots & 0 \\
 \vdots  & \vdots   & \ddots & \vdots \\
 0       & 0        & \cdots & g^{(L)}
\end{array}
\right)
\end{array}.
\] \\
QN-BDA approximation involves $L$ quantum evaluations and works well on models with weak correlation.

\textit{Quantum Natural-Simultaneous Perturbation Stochastic Approximation (QN-SPSA).}
Inheriting the SPSA idea to calculate the gradient, it is capable to put up generating the Hessian matrix with fewer evaluations, the so-called 2-SPSA. Normally, the HES or FIM matrix consumes $\mathcal{O}(p^2)$ quantum expectation. QN-BDA enables us to turn to relying on the number of layers $\mathcal{O}(L)$, saving us lots of computational resources. Even further, QN-SPSA just only executes four quantum runs to obtain the full second-order derivative matrix. Let's consider the Hessian of the Fubini-Study metric,
\begin{align}
     H_{ij}({\boldsymbol\theta}) \equiv g_{ij}({\boldsymbol\theta}) = - \left.\frac{1}{2}\partial_i\partial_j |\langle \psi_\theta , \psi_{\tilde{\theta}}\rangle|^2\right|_{\tilde{\theta}=\theta},
     \label{HesFu}
\end{align}
which is just the Hessian form of Eq.~(\ref{qngmatrix}) so that we can deploy the 2-SPSA method to generate the QN-SPSA matrix \cite{Gacon_2021,PhysRevA.103.012405}. We can see the equivalence,
\begin{align}
    &- \left.\frac{1}{2}\partial_i\partial_j |\langle \psi_\theta , \psi_{\tilde{\theta}}\rangle|^2\right|_{\tilde{\theta}=\theta} =\left. -\partial_i \text{Re}\left\{\langle \psi_\theta , \psi_{\tilde{\theta}}\rangle\langle \psi_{\tilde{\theta}},\partial_j\psi_\theta \rangle\right\}\right|_{\tilde{\theta}=\theta} \notag \\
    & = - \text{Re}\left\{-\langle \partial_i \psi_{\theta},  \partial_j\psi_\theta\rangle+\langle \partial_i\psi_\theta , \psi_{\theta}\rangle\langle \psi_{\theta}, \partial_j\psi_\theta \rangle\right\},
\end{align}
which is exactly the same as Eq.~(\ref{qngmatrix}).
Applying the 2-SPSA approach to compute the second-order derivative of the function $F({\boldsymbol\theta},\tilde{\theta}) = - \frac{1}{2}|\langle \psi_\theta , \psi_{\tilde{\theta}}\rangle|^2$ instead of our loss function $f({\boldsymbol\theta})$ as in Newton method. The core estimator in second-order SPSA is perturbed by two random vectors $\vec{\Delta}^1_k$, $\vec{\Delta}^2_k  \in \mathcal{U}^p\{-1,1\}$ at step $k$-th

\begin{equation}
\begin{aligned}
  \triangle F = \frac{-1}{2} &\left[F({\boldsymbol\theta}_k+s_k\vec{\Delta}^1_k+s_k\vec{\Delta}^2_k ,{\boldsymbol\theta}_k)
  \right.\\ &-F({\boldsymbol\theta}_k +s_k\vec{\Delta}^1_k ,{\boldsymbol\theta}_k)\\&+F({\boldsymbol\theta}_k-s_k\vec{\Delta}^1_k ,{\boldsymbol\theta}_k)
  \\ &\left.-F({\boldsymbol\theta}_k-s_k\vec{\Delta}^1_k+s_k\vec{\Delta}^2_k ,{\boldsymbol\theta}_k)\right],
\end{aligned}
\end{equation}
which is composed of four terms, corresponding to four quantum expectations we run on the quantum processor. Then, the Fubini-Study metric Eq.~(\ref{HesFu}) is replaced by the QN-SPSA metric at $k$-th iteration
\begin{equation}
    g^{k}({\boldsymbol\theta}) \rightarrow \bar{H}^{k}({\boldsymbol\theta}) = \frac{\triangle F}{4 s_k^2}  \left(\vec{\Delta}^1_k\vec{\Delta}^{2\text{T}}_k+\vec{\Delta}^2_k\vec{\Delta}^{1\text{T}}_k\right), 
\end{equation}
where $\bar{H}^{k}({\boldsymbol\theta}) \in \mathbb{R}^{p \times p}$ and $s_k$ is a small positive hyperparameter able to be tuned. The metric is, however, still too highly stochastic, so some helpful techniques are invoked to address this issue, for example by counting information from previous updates to smooth the estimator
\begin{equation}
    \tilde{H}^{k} = \frac{k}{k+1}\tilde{H}^{k-1}+\frac{1}{k+1}\bar{H}^{k},
\end{equation}
eventually, because the estimator is still ill-conditioned and unstable. To satisfy the positive semi-definite warranting the local convex analysis and invertibility condition of $\tilde{H}^{k}$, the following replacement is needed accordingly,
\begin{equation}
    \tilde{H}^{k} \rightarrow \sqrt{\tilde{H}^{k} \tilde{H}^{k} } +\beta \mathbb{1},
\end{equation}
where the second term $\beta \mathbb{1} \in \mathbb{R}^{p \times p}$ corresponds to the invertibility condition. The effect from the geometry metric is suppressed as the positive regulator gets a huge value $\beta \gg 0$, which indeed reduces to the standard gradient descent scenario and is more unstable. Therefore, that, in turn, causes a larger deviation in the average sample result when $\beta \rightarrow 0$. That makes a constant regulator $\beta$, which is thus a trade-off between quantum-geometric information and numerical instability. Another regularization approach, called “half-inversion”, generalizes the power of the second-order derivative matrix to $n$ instead of the usual ones $-1$ corresponding to standard natural gradient and $0$ standing for original gradient, given by \cite{haug2021optimal}.
 
\textit{Quantum Natural-Simultaneous Perturbation Stochastic Approximation with Parameter-Shift Rule (QN-SPSA+PSR)} is our extension for the QN-SPSA+SPSA method. Although QN-SPSA+SPSA has smaller complexity when compared with other methods, this stochastic method is highly stochastic and makes it harder to converge and less stable when it comes to a bigger cost function landscape. Here, we propose the QN-SPSA+PSR algorithm, which is inherent in approximating the Fubini-Study metric properties in a way of saving computational cost. However, instead of using the SPSA method to approximate the gradient of the cost function, we use the analytical method to calculate the gradient by PSR, which would compensate for the instability of the stochastic approximation in the Fubini-Study metric and drive itself in the gradient direction. QN-SPSA+PSR, therefore, as a practical hybrid variant within the broader quantum natural-gradient optimization family, which is not the algebraic combination itself, but the empirical observation that replacing the stochastic SPSA gradient with PSR stabilizes QN-SPSA-based updates in the TIM VQE benchmark, while preserving the low metric-estimation cost of QN-SPSA. This is also expected for more stable and faster convergence compared to the former QN-SPSA+SPSA method.
\section{\label{rad}Results and Discussion}
\subsection{\label{re}Complexity estimation}

\begin{table}[h]
\resizebox{0.5\textwidth}{!}{
    \colorbox{lightgray}{%
\begin{tabular}{|cccc|}
\hline
\rowcolor{blue!15}
\multicolumn{4}{|c|}{Derivative-based method efficiency}                                                                \\ \hline \hline
\multicolumn{1}{|c||}{$\triangledown f({\boldsymbol\theta})$}         & \multicolumn{1}{c|}{PSR}                   & \multicolumn{1}{c|}{FD}      & SPSA    \\ \hline
\rowcolor{LightCyan}
\multicolumn{1}{|c||}{Comp. cost}  & \multicolumn{1}{c|}{2p}                    & \multicolumn{1}{c|}{2p}      & 2       \\
\rowcolor{LightCyan}
\multicolumn{1}{|c||}{Approx.}  & \multicolumn{1}{c|}{Exact}                    & \multicolumn{1}{c|}{Finite Difference}      & Simultaneous Perturbation       \\ \hline \hline
\multicolumn{1}{|c||}{$g({\boldsymbol\theta})$} & \multicolumn{1}{c|}{QN-BDA}               & \multicolumn{1}{c|}{QN-SPSA} & Constant \\ \hline
\rowcolor{LightCyan}
\multicolumn{1}{|c||}{Comp. cost}  & \multicolumn{1}{c|}{L} & \multicolumn{1}{c|}{4}       & 0       \\
\rowcolor{LightCyan}
\multicolumn{1}{|c||}{Approx. }  & \multicolumn{1}{c|}{Block exact} & \multicolumn{1}{c|}{Simultaneous Perturbation}       & Identity       \\ \hline
\end{tabular}}}
\caption{Comparison of derivative-based optimizer components in terms of computational cost and approximation accuracy. The table summarizes the gradient-estimation methods and adaptive learning-rate schemes used in this work, together with their corresponding quantum-evaluation costs. The methods are ordered from left to right according to their closeness to the exact target quantity. For gradient estimation, the Parameter-Shift Rule (PSR) provides an exact analytic derivative for the considered parametrized quantum gates, Finite Difference (FD) introduces a finite-shift approximation error, and Simultaneous Perturbation Stochastic Approximation (SPSA) gives a stochastic gradient estimate using simultaneous random perturbations. For adaptive learning rate or metric-based methods, Quantum Natural Gradient with Block-Diagonal Approximation (QN-BDA) approximates the Fubini–Study metric by its block-diagonal structure, Quantum Natural Gradient with SPSA (QN-SPSA) estimates the metric stochastically, and the constant learning rate method corresponds to using the identity matrix. Here, p denotes the number of variational parameters, and L denotes the number of PQC layers sharing the same rotation generator.}
    \label{Comparision method}   
\end{table}


\begin{figure*}[htb!]
  \centering
  \includegraphics[width=\textwidth]{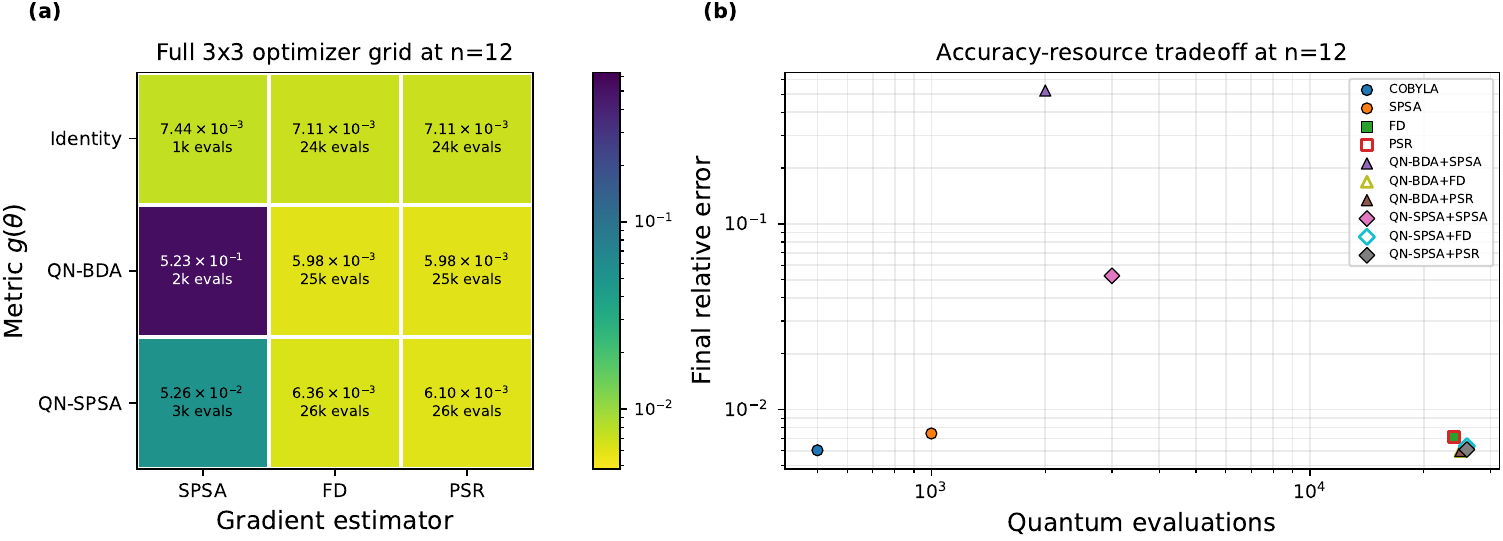}
  \caption{Numerical summary of the optimizer-comparison grid for the 12-qubit TIM benchmark with the two-layer linear entanglement RealAmplitudes ansatz, $J=1$, $h=2.0$, and 500 parameter updates. Left: complete $3 \times 3$ derivative-based grid, where each cell reports the final relative error and the total quantum evaluation count. Right: final relative error versus total quantum evaluations for the same benchmark, including COBYLA as a derivative-free baseline.}
  \label{error_qevl_method_comparision}
\end{figure*}

In summary of the derivatives-based optimization framework, we tell apart two subroutines: gradient calculation and $g({\boldsymbol\theta})$ matrix evaluation. For gradient calculation, we employ three approaches: PSR offering exact gradients of quantum ansatz structure, FD-a classical numerical method that approximates gradients, and SPSA-a stochastic approximation technique requiring only two function evaluations per iteration. For the $g({\boldsymbol\theta})$ matrix calculation, we generally have three metric choices: Constant-an identity matrix; QN-BDA approximates the Fubini-Study metric using the block diagonal structure; and QN-SPSA, which stochastically approximates the Fisher Information Matrix with only four evaluations per iteration. A method is a combination of the gradient and $g({\boldsymbol\theta})$, such as QN-BDA+PSR, which implies the application of the QN-BDA technique and PSR for the estimation of the gradient.

Tab.~\ref{Comparision method} summarizes the component costs and approximation types, demonstrating the trade-offs in performance and resource requirements among methods, e.g., accuracy, update speed, and computational cost. Tab. \ref{Comparision method} is derived from the underlying theories of methods discussed above. Besides that, method's orders in Tab. \ref{Comparision method} based on theory-driven and bias-informed conjectures from previous studies \cite{spall1998overview, Schuld_2019,Stokes_2020,Gacon_2021,PhysRevA.103.012405}, in which the methods such as, QN-BDA, QN-SPSA, and Constant, are ranked in descending order of convergence, accuracy, and computational cost. Similarly, for gradient methods, SPSA, FD, and PSR follow an ascending order of accuracy, convergence, and computational cost. Consequently, one can infer, for instance, that while the fully computed Hessian-like matrix+PSR should theoretically achieve the best performance at the expense of the intractably highest computational cost, the feasible approximate QN-BDA+PSR consumes much quantum overhead but is expected to outperform QN-SPSA+SPSA overall.
    
    
\begin{figure*}
    \centering
    \includegraphics[width=0.98\linewidth]{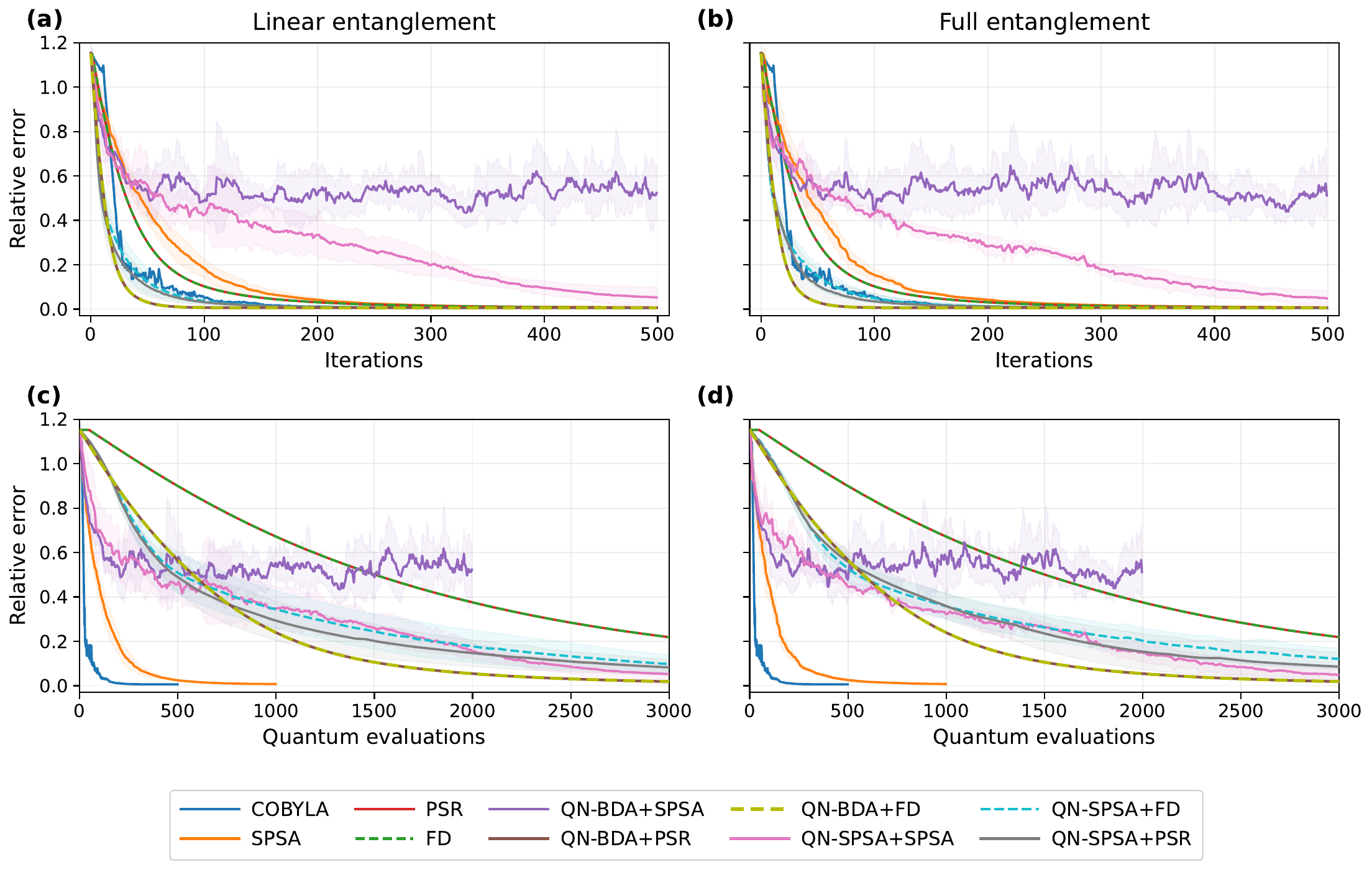}
    \caption{Comparison of optimization methods for the TIM model with 12 spins 
    using the RealAmplitudes ansatz. The relative error is shown as a function of 
    iterations (a,b) and quantum evaluations (c,d) for two 
    entanglement mappings: linear  (a,c) and full (b,d).
    }
    \label{method comparision}
\end{figure*}

\begin{table*}[t]
\centering
\resizebox{\textwidth}{!}{%
\begin{tabular}{lllcc}
\toprule
Optimizer & Metric & Gradient &  Relative error & Quantum evaluations \\
\midrule
COBYLA & -- & -- &  $6.04 \times 10^{-3}$ & 500 \\
SPSA & Identity & SPSA &  $7.44 \times 10^{-3} \pm 5.92 \times 10^{-4}$ & 1000 \\
FD & Identity & FD &  $7.11 \times 10^{-3}$ & 24000 \\
PSR & Identity & PSR & $7.11 \times 10^{-3}$ & 24000 \\
QN-BDA+SPSA & QN-BDA & SPSA &  $5.23 \times 10^{-1} \pm 1.25 \times 10^{-1}$ & 2000 \\
QN-BDA+FD & QN-BDA & FD &  $5.98 \times 10^{-3}$ & 25000 \\
QN-BDA+PSR & QN-BDA & PSR &  $5.98 \times 10^{-3} \pm 0$ & 25000 \\
QN-SPSA+SPSA & QN-SPSA & SPSA & $5.26 \times 10^{-2} \pm 4.68 \times 10^{-2}$ & 3000 \\
QN-SPSA+FD & QN-SPSA & FD &  $6.36 \times 10^{-3} \pm 5.86 \times 10^{-4}$ & 26000 \\
QN-SPSA+PSR & QN-SPSA & PSR & $6.10 \times 10^{-3} \pm 3.07 \times 10^{-4}$ & 26000 \\
\bottomrule
\end{tabular}}
\caption{Numerical accuracy summary for the 12-qubit optimizer comparison. The table reports final relative errors and the total quantum evaluation count after 500 parameter updates for the two-layer RealAmplitudes ansatz with linear entanglement, $J=1$, and $h=2.0$. For stochastic methods, values are reported as mean $\pm$ one standard deviation over seven random seeds, while deterministic methods have one run. In this setup, $p=24$ and $L=2$, giving per-update quantum costs of 1 for COBYLA, 2 for SPSA, $2p=48$ for FD/PSR, $L+2=4$ for QN-BDA+SPSA, $L+2p=50$ for QN-BDA+FD/PSR, 6 for QN-SPSA+SPSA, and $4+2p=52$ for QN-SPSA+FD/PSR.}
\label{tab:accuracy_summary}
\end{table*}

From this framework, a key motivation of the QN-SPSA+PSR algorithm is to merge the computational resource efficiency of QN-SPSA with the precision of the PSR. The computational cost of QN-SPSA, at four quantum evaluations per iteration, demonstrates remarkable scalability compared to traditional Hessian-based approaches like QN-BDA, which scale quadratically or linearly with the number of parameters and layers. While the PSR would keep the method at good stability and convergence. This refinement enables faster convergence and greater stability while maintaining a low computational overhead, highlighting its potential for practical application on NISQ devices. Notably, these are only theory-based rough estimations. In practice, the complexity of objective functions and the influence of noise would lead to variations in the performance of these methods, and approximate methods, such as methods using SPSA, would have advantages as noise comes about. The numerical results in this study, however, demonstrate an even better performance of QN-SPSA+PSR than predicted by the estimation, as we will present in the next section. For visual clarity, Fig.~\ref{method comparision} shows the convergence of representative methods, while the numerical summaries are reported in Fig.~\ref{error_qevl_method_comparision} and Tab.~\ref{tab:accuracy_summary} for the RealAmplitude ansatz with two-layer linear entanglement. We therefore avoid treating QN-SPSA+PSR as automatically superior from the component design alone, instead, the following section reports its measured accuracy next to all other implemented combinations and then compares convergence both by update count and by quantum-evaluation count.

\subsection{Simulation results}
\label{Simulation}

We performed a detailed investigation of the VQE algorithm applied to the TIM model, testing several prominent optimizers, ansatz structures, and entanglement configurations. Our simulations revealed several important insights into the performance of optimization methods and ansatz structures, highlighting the practical advantages of the proposed QN-SPSA+PSR algorithm.
Unless otherwise specified, the external field and coupling constants were set to the default values $h = 2$ and $J=1$, respectively. For the stochastic methods, which are evaluated based on 7 samples, and the blocking condition is applied to the parameter update \cite{spall1997accelerated}. Additionally, 500 iterations, two-layer linear entanglement mapping, learning rate = 0.01, all angles are initiated at 0.5, and a statevector simulator is used. For further information and input, the source code can be found at GitHub \cite{github}.

We ran the numerical study of the TIM model ($J=1$, and $h=2.0$) with 12 spins using the two-layer RealAmplitudes ansatz with linear entanglement, which are described in Fig.~\ref{error_qevl_method_comparision} and Tab.~\ref{tab:accuracy_summary}. From the information, we see that COBYLA is the cheapest one and reaches a comparable magnitude order of relative error with other methods. The QN-BDA+FD and QN-BDA+PSR rows
both reach the same error, showing that the FD and PSR
gradient choices give the same result for this metric in the present ansatz. Similarly, QN-SPSA+FD and QN-
SPSA+PSR agree within the stochastic spread, whereas QN-SPSA+SPSA and QN-BDA+SPSA combinations are cheaper but less stable and less accurate in this 12-qubit setting.
In Fig. \ref{method comparision}, we use the two-layer RealAmplitudes ansatz, linear and full entanglement, to compare ten different optimizers with nine combinations of gradient-based optimizers and COBYLA. The result is not significantly different between the two types of entanglement mapping. Remarkably, we observed in Fig. \ref{method comparision}(a,b) that QN-SPSA+PSR converges faster than most methods per parameter update, improving substantially over QN-SPSA+SPSA in this benchmark and reaching a final accuracy comparable to QN-BDA+PSR. While the QN-BDA+PSR method behaves as expected, as predicted by theory, showing the fastest convergence by direct use of the Fubini-Study metric tensor, QN-SPSA+PSR achieved comparable results with lower computational overhead. By a small change in the gradient part from stochastic to exact estimation, we improved the performance of the inspired QN-SPSA+SPSA method, which made QN-SPSA+PSR more stable and faster to converge. Additionally, the classical derivative-free COBYLA exhibits a striking achievement, measuring up to QN-BDA+PSR and surpassing several algorithms while needing the least quantum overhead of cost function evaluation, one per iteration. That makes it an appealing method to implement in the hybrid algorithm. In another aspect, we present them in Fig. \ref{method comparision}(c, d), which illustrates the practical computational cost of each method, measured in terms of the number of quantum evaluations required at each parameter-update step. Since different optimization methods require different numbers of quantum evaluations per iteration, the overall computational effort varies accordingly, as these quantum calls contribute directly to the execution time. For the 12-qubit RealAmplitudes ansatz with two layers, the number of quantum evaluations per iteration is as follows: SPSA requires 2 evaluations, COBYLA requires 1, FD/PSR requires 48, QN-BDA+PSR/FD requires 50, QN-SPSA+PSR/FD requires 52, QN-SPSA+SPSA requires 6, and QN-BDA+SPSA needs 4. Therefore, COBYLA and SPSA provide efficient and practical optimization strategies for hybrid applications on near-term quantum devices. In this simple model, QN-BDA+PSR, QN-SPSA+SPSA, and QN-SPSA+PSR exhibit slower performance and are broadly comparable to one another. However, for more complex problem structures, leveraging quantum-state information may accelerate convergence to the global minimum within a feasible timescale, particularly in regimes where classical methods become inefficient. In this context, access to an accurate Quantum Fisher Information Matrix is especially advantageous, and since the computation cost does not scale exponentially in number of qubits, it opens up a way to collaborate with HPC workflow to leverage the full potential of these quantum optimizers.

\begin{figure} [htp!]
    \centering
\includegraphics[scale=0.5]{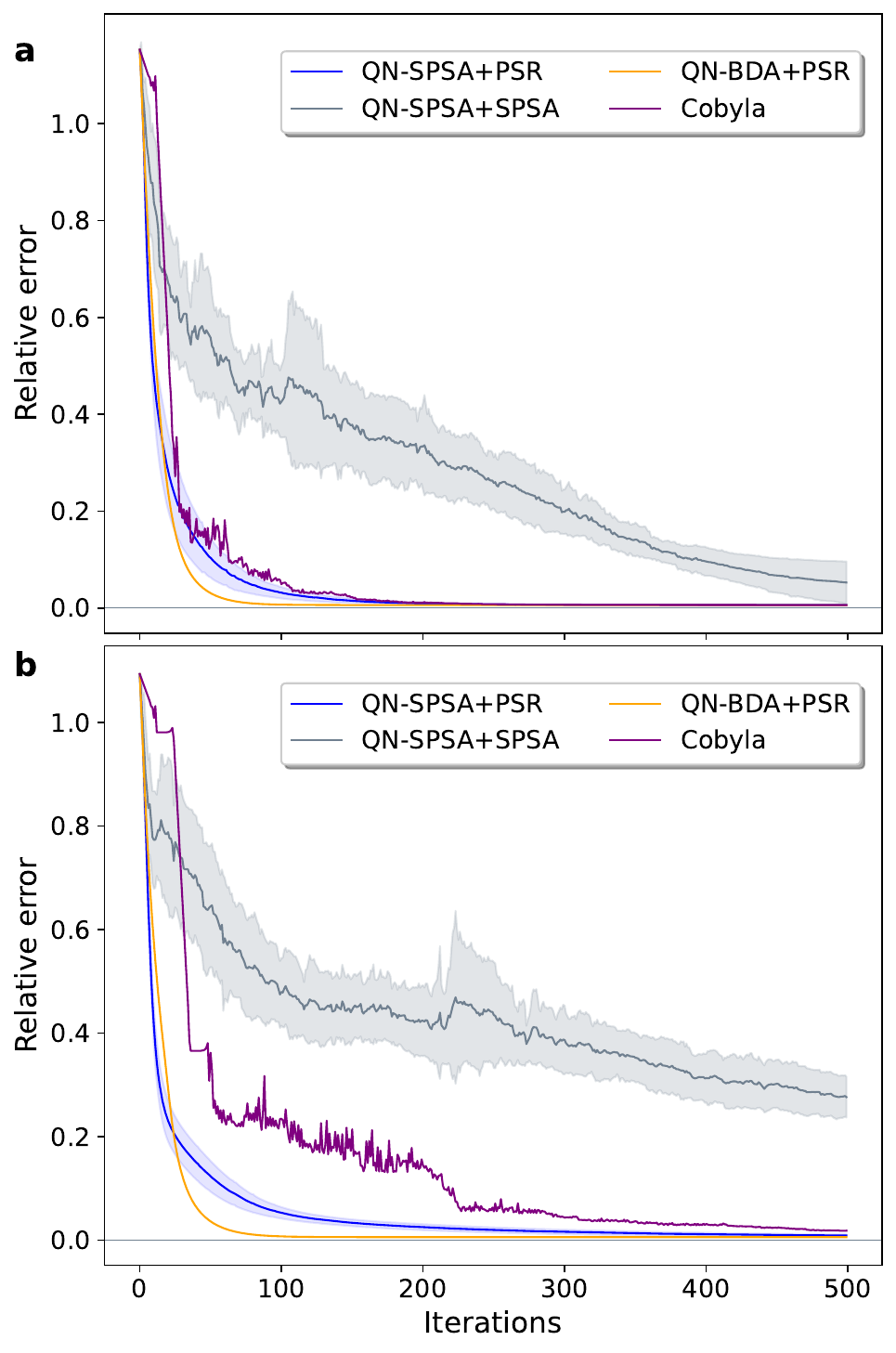}
    \caption{Comparison of optimization methods for the TIM model with 12 spins using different ansatz configurations. The plots illustrate the relative error as a function of iterations for multiple optimization strategies applied to (a) the RealAmplitudes ansatz and (b) the EfficientSU2 ansatz. The visualized outcomes highlight that the RealAmplitudes ansatz achieves better and more consistent performance than the EfficientSU2 ansatz for the TIM problem, as previously discussed in Section \ref{Ansatz selection}.}
    \label{ansatz compare}
\end{figure}

\begin{figure} [htp!]
    \centering
\includegraphics[scale=0.5]{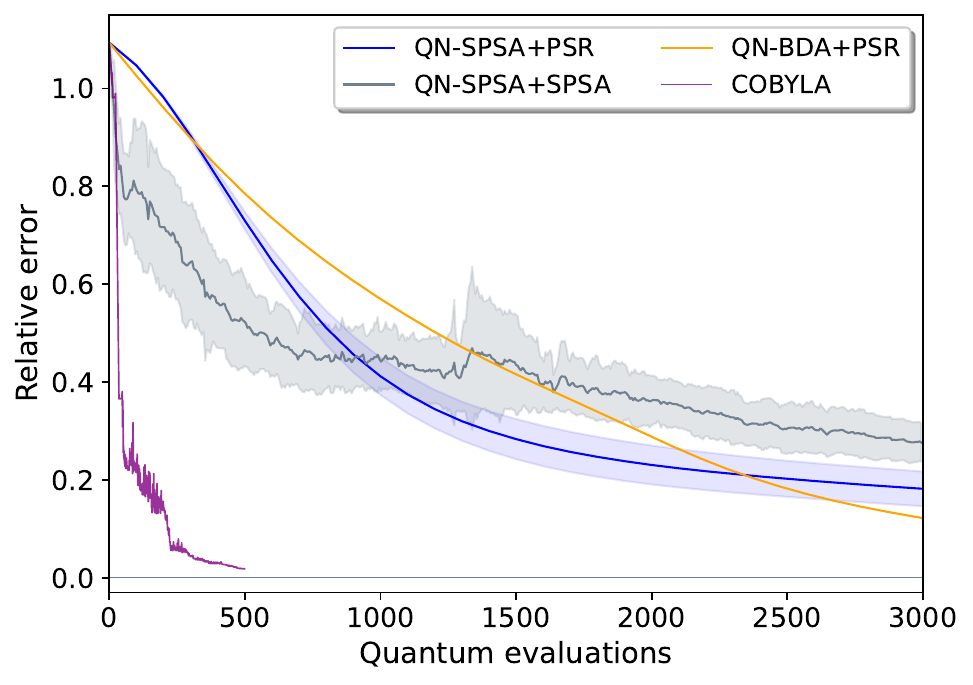
}
    \caption{The relative error is plotted as a function of the number of quantum evaluations for several optimization strategies using the EfficientSU2 ansatz. For the 12-qubit EfficientSU2 ansatz with four layers, the number of quantum evaluations per iteration is SPSA: 2, QN-BDA+PSR: 100, FD: 96, COBYLA: 1, QN-SPSA+PSR: 100, QN-SPSA+SPSA: 6.}
    \label{ansatz_compare_quantum_eval}
\end{figure}

\begin{figure}[htp]
    \centering

    \begin{subfigure}{0.45\textwidth}
        \centering
        \includegraphics[width=\textwidth]{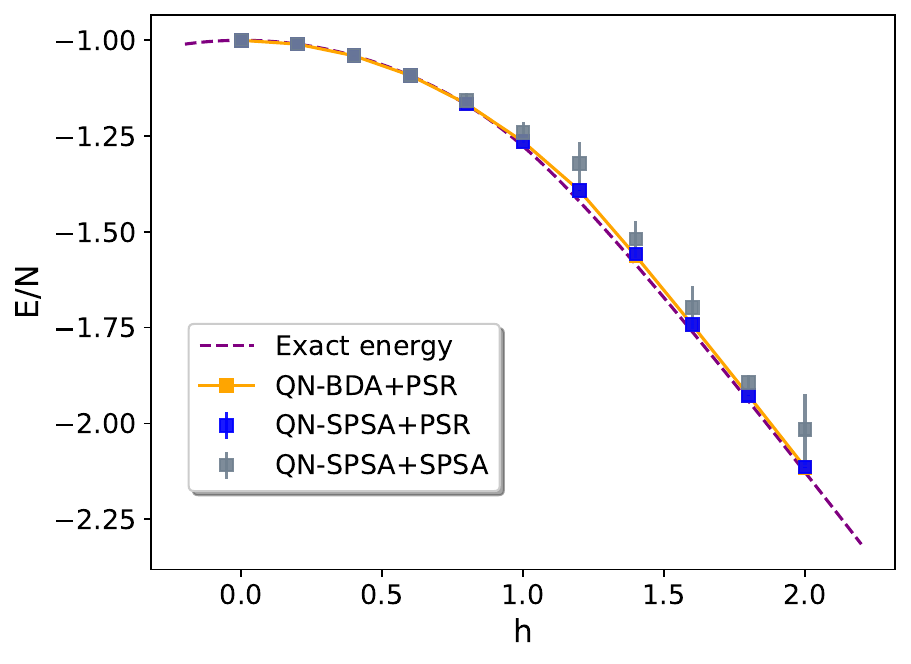}
        \caption*{}
    \end{subfigure}
    \begin{subfigure}{0.45\textwidth}
        \centering
        \includegraphics[width=\textwidth]{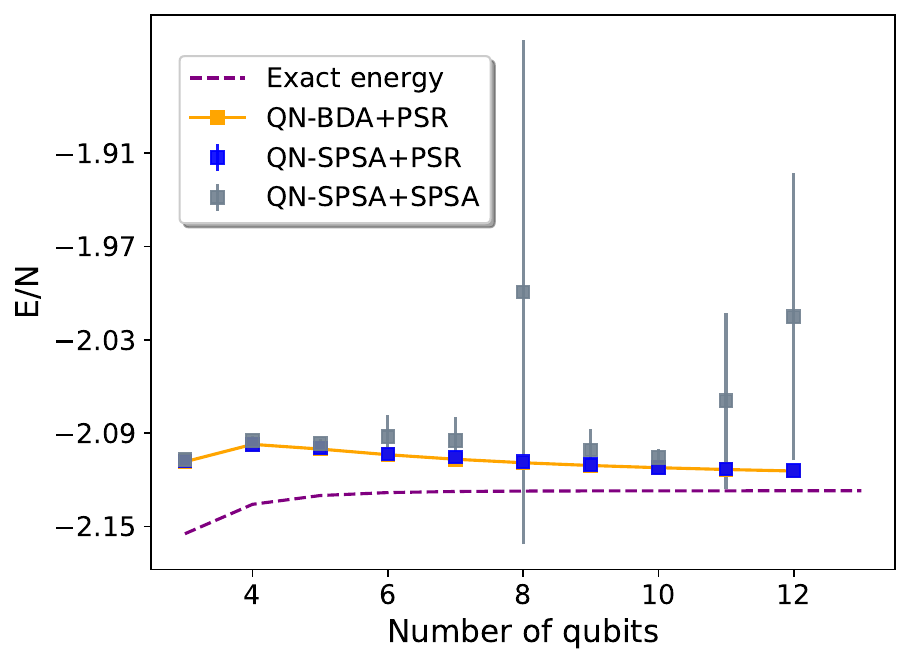}
        \caption*{}
    \end{subfigure}

    \caption{
Estimated average ground state energy for TIM using
QN-SPSA+SPSA, QN-SPSA+PSR, and QN-BDA+PSR with the RealAmplitudes ansatz.
The upper panel shows results for a 12-spin system under different external field strengths,
while the lower panel shows results for different system sizes.
}
    \label{fig:8}
\end{figure}

 In the simple TIM setup and ansatz structure considered here, FD and PSR exhibit similar optimization trajectories and get used the same computational overheads. Therefore, we will treat them alternatively in the combination methods presented in the following results. Another typical combination is QN-BDA+SPSA, where the quantum natural gradient is estimated using the block-diagonal approximation, while the gradient is estimated via simultaneous perturbative approximation, leading to one of the resource-efficient methods. However, this method is not considered in the following because its performance differs substantially from that of the other methods. An additional  investigation of this case is provided in Appendix.
Next, we focus primarily on comparing four methods that are variants of the quantum natural gradient approach, which constitute the main methods of interest in this study, together with COBYLA, which is included due to its relatively low computational cost. Similar to Fig.~\ref{method comparision}, Fig.~\ref{ansatz compare} illustrates their convergence behavior in terms of information updates and demonstrates their compatibility with different ansatz structures. The improved convergence observed for the RealAmplitudes ansatz compared with EfficientSU2 supports the suitability of the chosen ansatz, as discussed in Sec.~\ref{Ansatz selection}. For the EfficientSU2 ansatz, Fig.~\ref{ansatz_compare_quantum_eval} further highlights the advantage of COBYLA in terms of its low computational cost. However, the slower convergence of COBYLA shown in Fig.~\ref{ansatz compare} indicates the limitations of classical optimization methods with no access to quantum-state information when applied to more complex ansatz structures, as discussed earlier. For the EfficientSU2 ansatz, it should be noted that the layer index $L$ used in the QN-BDA computation is not equivalent to the ansatz layers. In this case, the relation is given by $L = 2 \times (\#\text{ansatz layers})$, since each ansatz layer includes two rotation gate layers of $R_Y$ and $R_Z$. Finally, Fig.~\ref{fig:8} presents estimates of the average ground state energy for various system sizes and external field strengths. Therein, after 500 iterations, the QN-SPSA+PSR and QN-BDA+PSR come relatively close to the exact result, meanwhile the QN-SPSA+SPSA has a wide standard deviation.

\begin{figure*}[htp!]
    \centering

    \begin{subfigure}[b]{\textwidth}
        \centering
        \includegraphics[width=\textwidth]{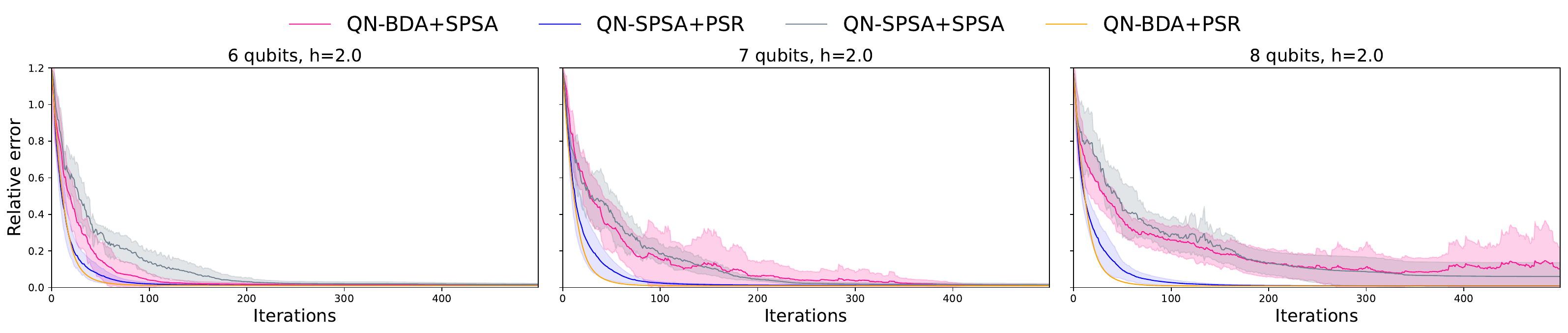}
        \caption*{}
    \end{subfigure}

    \begin{subfigure}[b]{\textwidth}
        \centering
        \includegraphics[width=\textwidth]{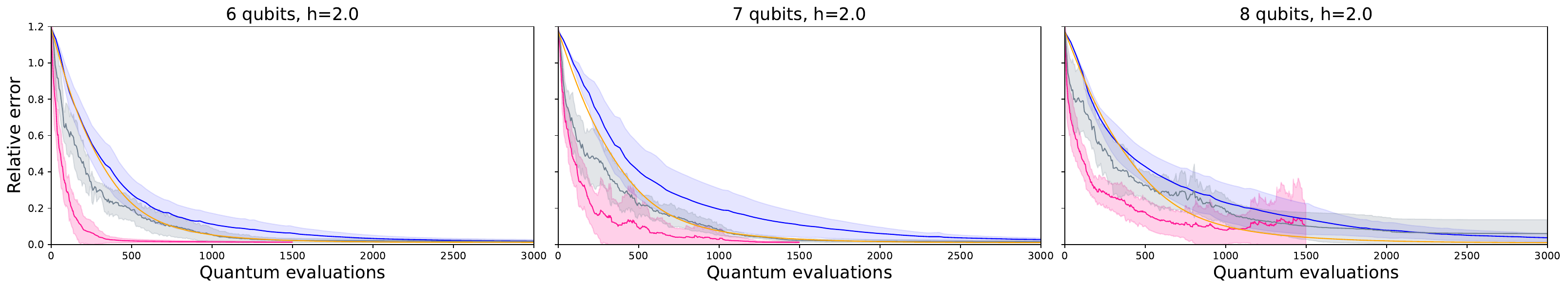}
        \caption*{}
    \end{subfigure}

    \caption{The plots show the relative error of four typical optimizer combinations as a function of parameter updating step (upper plots) and quantum evaluations (lower plots). The results run for the TIM model with (6,7,8) qubits using the RealAmplitudes ansatz applied with two-layer linear entanglement mapping as in Fig.~\ref{realanstz}a.}
    \label{fig:smallqubit_combinations_optimization_survey}
\end{figure*}

\begin{figure*}[htp!]
    \centering

    \begin{subfigure}[b]{\textwidth}
        \centering
        \includegraphics[width=\textwidth]{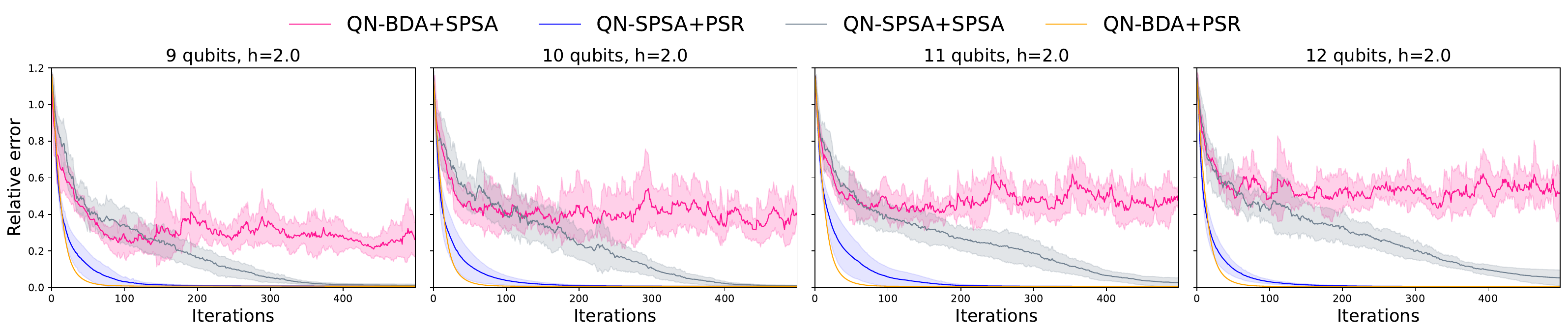}
        \caption*{}
    \end{subfigure}

    \begin{subfigure}[b]{\textwidth}
        \centering
        \includegraphics[width=\textwidth]{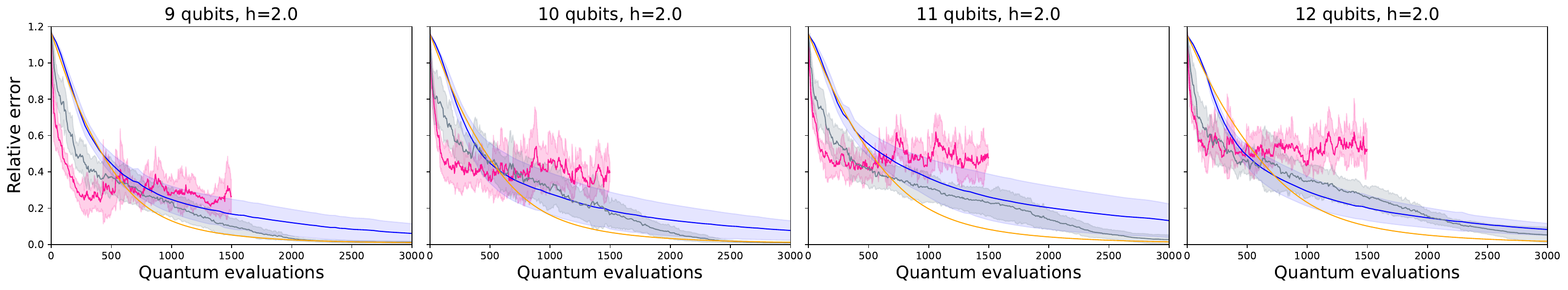}
        \caption*{}
    \end{subfigure}

    \caption{The plots show the relative error of four typical optimizer combinations as a function of parameter updating step (upper plots) and quantum evaluations (lower plots). The results run for the TIM model with (9,10,11,12) qubits using the RealAmplitudes ansatz applied with two-layer linear entanglement mapping as in Fig.~\ref{realanstz}a.}
    \label{fig:combinations_optimization_survey}
\end{figure*}

\section{Conclusion}
\label{conclusion}
The robustness of the QN-SPSA+PSR method further manifests using the more complex EfficientSU2 ansatz, which introduces a highly intricate cost function, shown in Fig. \ref{ansatz compare}. Although the EfficientSU2 ansatz would create a more complex cost function, QN-SPSA+PSR still shows stable and fast convergence properties and achieves results comparable to QN-BDA+PSR optimizer. Whereas, QN-SPSA+SPSA and COBYLA showed signs of stagnation and requires more iterations to approach optimal solutions.
Furthermore, Figs. \ref{method comparision} and \ref{ansatz compare}, which compared two types of ansatz and two types of entanglement schemes, show that linear entanglement is not much different than the full entanglement, and the RealAmplitudes performed better in convergence than Efficient SU2. That proved the initial arguments aforementioned in Section \ref{Ansatz selection}, the RealAmplitudes with linear entanglement, therefore, is sufficient for the TIM benchmarks considered here. On the other hand, Figs.~\ref{ansatz_compare_quantum_eval} and
\ref{method comparision}(c, d) demonstrate the practical advantage of
COBYLA and SPSA-based methods for near-term applications, since they require very low computational cost. To this end, it should not be
interpreted as evidence of the downside of the quantum methods in practice. The present study is based on a relatively simple model with weak
quantum correlations, where classical or stochastic methods can still perform
efficiently.
From the perspective of quantum information, the rapid convergence of the
quantum-informed update direction indicate that quantum-informed metric updates can improve stability for some optimizer combinations, where non-quantum-informed optimizers may
struggle to reach the global minimum. These observations motivate further investigation of quantum optimizers in larger and more strongly correlated systems. Finally, the TIM energy estimates are obtained by varying external fields and numbers of spins, as illustrated in Figs. \ref{fig:8}. In both cases, the QN-SPSA+PSR consistently delivered reliable and accurate results for the average ground state energy, closely aligning with the QN-BDA+PSR.

In conclusion, this work provides a comprehensive accuracy-resource review and comparison of classical, stochastic, and quantum-informed optimizers. Moreover, leveraging the model's properties and symmetries, we analyze the Ising Model to inform a physically motivated ansatz.

Our numerical results indicated that the chosen ansatz form, based on the Hamiltonian's properties, performed reasonably well. Although estimating the number of layers in Eq. \ref{layer estimation} not necessarily being employed in this case, it is expected to serve as a reasonably good initial estimation as the system scales up.

Our exploration of optimization methods
has yielded encouraging outcomes. The newly proposed QN-SPSA+PSR algorithm exhibits highly favorable computational efficiency. It demonstrates potential not only in reducing computational overhead, since it does not scale up significantly with qubits like traditional Quantum Natural Gradient methods, but also in achieving more accurate optimal solutions. This work suggests promising prospects for further exploration and application of such algorithms for future research endeavors in QML, a wide range of VQAs, and NISQ devices.

\section*{Acknowledgments}

The authors acknowledge helpful discussions and contributions from Triet Minh Ha. The numerical calculations in this work are performed using the high-performance computing systems at Phenikaa University and PGI-12, Forschungszentrum Jülich.\\
\textbf{Author contributions} V.D. Nguyen and H.Q. Nguyen guided and supervised the project. The ideas and results were developed and carried out by D.T. Le and V.L. Nguyen. D.T. Le wrote the manuscript with assistance from V.L. Nguyen and feedback from  H.Q. Nguyen, V.D. Nguyen, and H.C. Nguyen. V.L. Nguyen packaged the code and executed the final summarization results on the HPC systems. All authors read and approved the final manuscript.\\
\textbf{Data Availability} The datasets generated and analysed during the current study are available in the Variational-Quantum-EigeinSolver repository, \\ \href{https://github.com/nguyenvulinh666/Variational-Quantum-EigeinSolver}{https://github.com/nguyenvulinh666/Variational-Quantum-EigeinSolver}.\\
\textbf{List of abbreviations} VQE, NISQ, VQA, QPE, QAOA, PQC, TIM, QML, GD, COBYLA, FD, PSR, SPSA, QNG, HES, FIM, QGT, QN-BDA, QN-SPSA.\\
\textbf{Competing interests}
The authors declare no competing interests.
 \section*{Appendix}

\label{Appendix}

The following investigation considers different combinations of  $g({\boldsymbol\theta})$ and $\triangledown f({\boldsymbol\theta})$, focusing on the four representative cases of Quantum Fisher Information Matrix, says, QN-BDA+SPSA, QN-SPSA+PSR, QN-SPSA+SPSA, QN-BDA+PSR. For the simple TIM model considered here, FD and PSR exhibit comparable performance. Therefore, we select QN-BDA+PSR and QN-BDA+SPSA as two representative QN-BDA-based approaches, and QN-SPSA+PSR and QN-SPSA+SPSA as two typical QN-SPSA-based cases. Although QN-BDA+SPSA is resource-efficient and performs well for small numbers of qubits presented in Fig. \ref{fig:smallqubit_combinations_optimization_survey}, its performance deteriorates with increasing system size as shown in Fig.~\ref{fig:combinations_optimization_survey}. Consequently, we have excluded QN-BDA+SPSA from the main results.

\EOD


\begin{thebibliography}{00}

\bibitem{cerezo2021variational}
Marco Cerezo, Andrew Arrasmith, Ryan Babbush, Simon~C Benjamin, Suguru Endo,
  Keisuke Fujii, Jarrod~R McClean, Kosuke Mitarai, Xiao Yuan, Lukasz Cincio,
  et~al.
\newblock Variational quantum algorithms.
\newblock {\em Nature Reviews Physics}, 3(9):625--644, 2021.

\bibitem{PhysRevA.92.042303}
Dave Wecker, Matthew~B. Hastings, and Matthias Troyer.
\newblock Progress towards practical quantum variational algorithms.
\newblock {\em Phys. Rev. A}, 92:042303, Oct 2015.

\bibitem{Preskill_2018}
John Preskill.
\newblock Quantum computing in the nisq era and beyond.
\newblock {\em Quantum}, 2:79, August 2018.

\bibitem{Brooks}
Michael Brooks.
\newblock Beyond quantum supremacy: the hunt for useful quantum computers.
\newblock {\em Nature}, 574:19--21, 10 2019.

\bibitem{Zhong_2020}
Han-Sen Zhong, Hui Wang, Yu-Hao Deng, Ming-Cheng Chen, Li-Chao Peng, Yi-Han
  Luo, Jian Qin, Dian Wu, Xing Ding, Yi~Hu, Peng Hu, Xiao-Yan Yang, Wei-Jun
  Zhang, Hao Li, Yuxuan Li, Xiao Jiang, Lin Gan, Guangwen Yang, Lixing You,
  Zhen Wang, Li~Li, Nai-Le Liu, Chao-Yang Lu, and Jian-Wei Pan.
\newblock Quantum computational advantage using photons.
\newblock {\em Science}, 370(6523):1460–1463, December 2020.

\bibitem{Arute:2019zxq}
Frank Arute et~al.
\newblock {Quantum supremacy using a programmable superconducting processor}.
\newblock {\em Nature}, 574(7779):505--510, 2019.

\bibitem{Wu_2021}
Yulin Wu, Wan-Su Bao, Sirui Cao, Fusheng Chen, Ming-Cheng Chen, Xiawei Chen,
  Tung-Hsun Chung, Hui Deng, Yajie Du, Daojin Fan, Ming Gong, Cheng Guo, Chu
  Guo, Shaojun Guo, Lianchen Han, Linyin Hong, He-Liang Huang, Yong-Heng Huo,
  Liping Li, Na~Li, Shaowei Li, Yuan Li, Futian Liang, Chun Lin, Jin Lin,
  Haoran Qian, Dan Qiao, Hao Rong, Hong Su, Lihua Sun, Liangyuan Wang, Shiyu
  Wang, Dachao Wu, Yu~Xu, Kai Yan, Weifeng Yang, Yang Yang, Yangsen Ye,
  Jianghan Yin, Chong Ying, Jiale Yu, Chen Zha, Cha Zhang, Haibin Zhang, Kaili
  Zhang, Yiming Zhang, Han Zhao, Youwei Zhao, Liang Zhou, Qingling Zhu,
  Chao-Yang Lu, Cheng-Zhi Peng, Xiaobo Zhu, and Jian-Wei Pan.
\newblock Strong quantum computational advantage using a superconducting
  quantum processor.
\newblock {\em Physical Review Letters}, 127(18), October 2021.

\bibitem{arute2019quantum}
Frank Arute, Kunal Arya, Ryan Babbush, Dave Bacon, Joseph~C Bardin, Rami
  Barends, Rupak Biswas, Sergio Boixo, Fernando~GSL Brandao, David~A Buell,
  et~al.
\newblock Quantum supremacy using a programmable superconducting processor.
\newblock {\em Nature}, 574(7779):505--510, 2019.

\bibitem{bharti2021noisy}
Kishor Bharti, Alba Cervera-Lierta, Thi~Ha Kyaw, Tobias Haug, Sumner
  Alperin-Lea, Abhinav Anand, Matthias Degroote, Hermanni Heimonen, Jakob~S
  Kottmann, Tim Menke, et~al.
\newblock Noisy intermediate-scale quantum (nisq) algorithms.
\newblock {\em arXiv preprint arXiv:2101.08448}, 2021.

\bibitem{peruzzo2014variational}
Alberto Peruzzo, Jarrod McClean, Peter Shadbolt, Man-Hong Yung, Xiao-Qi Zhou,
  Peter~J Love, Al{\'a}n Aspuru-Guzik, and Jeremy~L O’brien.
\newblock A variational eigenvalue solver on a photonic quantum processor.
\newblock {\em Nature communications}, 5(1):1--7, 2014.

\bibitem{farhi2014quantumapproximateoptimizationalgorithm}
Edward Farhi, Jeffrey Goldstone, and Sam Gutmann.
\newblock A quantum approximate optimization algorithm, 2014.

\bibitem{Schuld_2014}
Maria Schuld, Ilya Sinayskiy, and Francesco Petruccione.
\newblock An introduction to quantum machine learning.
\newblock {\em Contemporary Physics}, 56(2):172--185, October 2014.

\bibitem{mcclean2016theory}
Jarrod~R McClean, Jonathan Romero, Ryan Babbush, and Al{\'a}n Aspuru-Guzik.
\newblock The theory of variational hybrid quantum-classical algorithms.
\newblock {\em New Journal of Physics}, 18(2):023023, 2016.

\bibitem{PhysRevA.95.042308}
Jarrod~R. McClean, Mollie~E. Kimchi-Schwartz, Jonathan Carter, and Wibe~A.
  de~Jong.
\newblock Hybrid quantum-classical hierarchy for mitigation of decoherence and
  determination of excited states.
\newblock {\em Phys. Rev. A}, 95:042308, Apr 2017.

\bibitem{RevModPhys.40.153}
D.~J. ROWE.
\newblock Equations-of-motion method and the extended shell model.
\newblock {\em Rev. Mod. Phys.}, 40:153--166, Jan 1968.

\bibitem{Higgott_2019}
Oscar Higgott, Daochen Wang, and Stephen Brierley.
\newblock Variational quantum computation of excited states.
\newblock {\em Quantum}, 3:156, July 2019.

\bibitem{PhysRevResearch.1.033062}
Ken~M. Nakanishi, Kosuke Mitarai, and Keisuke Fujii.
\newblock Subspace-search variational quantum eigensolver for excited states.
\newblock {\em Phys. Rev. Res.}, 1:033062, Oct 2019.

\bibitem{PhysRevApplied.11.044092}
M.~Ganzhorn, D.J. Egger, P.~Barkoutsos, P.~Ollitrault, G.~Salis, N.~Moll,
  M.~Roth, A.~Fuhrer, P.~Mueller, S.~Woerner, I.~Tavernelli, and S.~Filipp.
\newblock Gate-efficient simulation of molecular eigenstates on a quantum
  computer.
\newblock {\em Phys. Rev. Appl.}, 11:044092, Apr 2019.

\bibitem{doi:10.1021/acs.jpclett.3c03159}
C{\'e}sar Feniou, Olivier Adjoua, Baptiste Claudon, Julien Zylberman, Emmanuel
  Giner, and Jean-Philip Piquemal.
\newblock Sparse quantum state preparation for strongly correlated systems.
\newblock {\em The Journal of Physical Chemistry Letters}, 15(11):3197--3205,
  2024.
\newblock PMID: 38483286.

\bibitem{Cao_2022}
Changsu Cao, Jiaqi Hu, Wengang Zhang, Xusheng Xu, Dechin Chen, Fan Yu, Jun Li,
  Han-Shi Hu, Dingshun Lv, and Man-Hong Yung.
\newblock Progress toward larger molecular simulation on a quantum computer:
  Simulating a system with up to 28 qubits accelerated by point-group symmetry.
\newblock {\em Physical Review A}, 105(6), June 2022.

\bibitem{PhysRevX.6.031007}
P.~J.~J. O'Malley, R.~Babbush, I.~D. Kivlichan, J.~Romero, J.~R. McClean,
  R.~Barends, J.~Kelly, P.~Roushan, A.~Tranter, N.~Ding, B.~Campbell, Y.~Chen,
  Z.~Chen, B.~Chiaro, A.~Dunsworth, A.~G. Fowler, E.~Jeffrey, E.~Lucero,
  A.~Megrant, J.~Y. Mutus, M.~Neeley, C.~Neill, C.~Quintana, D.~Sank,
  A.~Vainsencher, J.~Wenner, T.~C. White, P.~V. Coveney, P.~J. Love, H.~Neven,
  A.~Aspuru-Guzik, and J.~M. Martinis.
\newblock Scalable quantum simulation of molecular energies.
\newblock {\em Phys. Rev. X}, 6:031007, Jul 2016.

\bibitem{PhysRevX.8.011021}
J.~I. Colless, V.~V. Ramasesh, D.~Dahlen, M.~S. Blok, M.~E. Kimchi-Schwartz,
  J.~R. McClean, J.~Carter, W.~A. de~Jong, and I.~Siddiqi.
\newblock Computation of molecular spectra on a quantum processor with an
  error-resilient algorithm.
\newblock {\em Phys. Rev. X}, 8:011021, Feb 2018.

\bibitem{Kandala_2017}
Abhinav Kandala, Antonio Mezzacapo, Kristan Temme, Maika Takita, Markus Brink,
  Jerry~M. Chow, and Jay~M. Gambetta.
\newblock Hardware-efficient variational quantum eigensolver for small
  molecules and quantum magnets.
\newblock {\em Nature}, 549(7671):242--246, sep 2017.

\bibitem{NielQuant}
Isaac L.~Chuang Michael A.~Nielsen.
\newblock {\em Quantum Computation and Quantum Information: 10th Anniversary
  Edition}.
\newblock Cambridge University Press, New York, 2010.

\bibitem{8585034}
Y.~Cao, J.~Romero, and A.~Aspuru-Guzik.
\newblock Potential of quantum computing for drug discovery.
\newblock {\em IBM Journal of Research and Development}, 62(6):6:1--6:20, 2018.

\bibitem{Blunt_2022}
Nick~S. Blunt, Joan Camps, Ophelia Crawford, Róbert Izsák, Sebastian
  Leontica, Arjun Mirani, Alexandra~E. Moylett, Sam~A. Scivier, Christoph
  Sünderhauf, Patrick Schopf, Jacob~M. Taylor, and Nicole Holzmann.
\newblock Perspective on the current state-of-the-art of quantum computing for
  drug discovery applications.
\newblock {\em Journal of Chemical Theory and Computation}, 18(12):7001–7023,
  November 2022.

\bibitem{Lordi_2021}
Vincenzo Lordi and John Nichol.
\newblock Advances and opportunities in materials science for scalable quantum
  computing.
\newblock {\em MRS Bulletin}, 46, 07 2021.

\bibitem{Cao_2019}
Yudong Cao, Jonathan Romero, Jonathan~P. Olson, Matthias Degroote, Peter~D.
  Johnson, Mária Kieferová, Ian~D. Kivlichan, Tim Menke, Borja Peropadre,
  Nicolas P.~D. Sawaya, Sukin Sim, Libor Veis, and Alán Aspuru-Guzik.
\newblock Quantum chemistry in the age of quantum computing.
\newblock {\em Chemical Reviews}, 119(19):10856–10915, August 2019.

\bibitem{Bauer_2020}
Bela Bauer, Sergey Bravyi, Mario Motta, and Garnet Kin-Lic Chan.
\newblock Quantum algorithms for quantum chemistry and quantum materials
  science.
\newblock {\em Chemical Reviews}, 120(22):12685–12717, October 2020.

\bibitem{Boixo_2018}
Sergio Boixo, Sergei~V. Isakov, Vadim~N. Smelyanskiy, Ryan Babbush, Nan Ding,
  Zhang Jiang, Michael~J. Bremner, John~M. Martinis, and Hartmut Neven.
\newblock Characterizing quantum supremacy in near-term devices.
\newblock {\em Nature Physics}, 14(6):595–600, April 2018.

\bibitem{mccaskey2019}
Alexander~J. McCaskey, Zachary~P. Parks, Jacek Jakowski, Shirley~V. Moore,
  T.~Morris, Travis~S. Humble, and Raphael~C. Pooser.
\newblock Quantum chemistry as a benchmark for near-term quantum computers,
  2019.

\bibitem{tilly2021variational}
Jules Tilly, Hongxiang Chen, Shuxiang Cao, Dario Picozzi, Kanav Setia, Ying Li,
  Edward Grant, Leonard Wossnig, Ivan Rungger, George~H Booth, et~al.
\newblock The variational quantum eigensolver: a review of methods and best
  practices.
\newblock {\em arXiv preprint arXiv:2111.05176}, 2021.

\bibitem{fedorov2021vqemethodshortsurvey}
Dmitry~A. Fedorov, Bo~Peng, Niranjan Govind, and Yuri Alexeev.
\newblock Vqe method: A short survey and recent developments, 2021.

\bibitem{Grant_2018}
Edward Grant, Marcello Benedetti, Shuxiang Cao, Andrew Hallam, Joshua Lockhart,
  Vid Stojevic, Andrew~G. Green, and Simone Severini.
\newblock Hierarchical quantum classifiers.
\newblock {\em npj Quantum Information}, 4(1), dec 2018.

\bibitem{doi:10.1126/science.aag2302}
Giuseppe Carleo and Matthias Troyer.
\newblock Solving the quantum many-body problem with artificial neural
  networks.
\newblock {\em Science}, 355(6325):602--606, 2017.

\bibitem{PhysRevX.10.031064}
Xavier Bonet-Monroig, Ryan Babbush, and Thomas~E. O'Brien.
\newblock Nearly optimal measurement scheduling for partial tomography of
  quantum states.
\newblock {\em Phys. Rev. X}, 10:031064, Sep 2020.

\bibitem{Wang_2019}
Daochen Wang, Oscar Higgott, and Stephen Brierley.
\newblock Accelerated variational quantum eigensolver.
\newblock {\em Physical Review Letters}, 122(14), apr 2019.

\bibitem{DEGENNES1963132}
P.G. {de Gennes}.
\newblock Collective motions of hydrogen bonds.
\newblock {\em Solid State Communications}, 1(6):132--137, 1963.

\bibitem{powell_1998}
M.~J.~D. Powell.
\newblock Direct search algorithms for optimization calculations.
\newblock {\em Acta Numerica}, 7:287–336, 1998.

\bibitem{powell2007view}
Michael~JD Powell.
\newblock A view of algorithms for optimization without derivatives.
\newblock {\em Mathematics Today-Bulletin of the Institute of Mathematics and
  its Applications}, 43(5):170--174, 2007.

\bibitem{conn1997convergence}
Andrew~R Conn, Katya Scheinberg, and Ph~L Toint.
\newblock On the convergence of derivative-free methods for unconstrained
  optimization.
\newblock {\em Approximation theory and optimization: tributes to MJD Powell},
  pages 83--108, 1997.

\bibitem{powell1994direct}
Michael~JD Powell.
\newblock A direct search optimization method that models the objective and
  constraint functions by linear interpolation.
\newblock In {\em Advances in optimization and numerical analysis}, pages
  51--67. Springer, 1994.

\bibitem{powell2002uobyqa}
Michael~JD Powell.
\newblock Uobyqa: unconstrained optimization by quadratic approximation.
\newblock {\em Mathematical Programming}, 92(3):555--582, 2002.

\bibitem{powell2006newuoa}
Michael~JD Powell.
\newblock The newuoa software for unconstrained optimization without
  derivatives.
\newblock In {\em Large-scale nonlinear optimization}, pages 255--297.
  Springer, 2006.

\bibitem{powell2009bobyqa}
Michael~JD Powell.
\newblock The bobyqa algorithm for bound constrained optimization without
  derivatives.
\newblock {\em Cambridge NA Report NA2009/06, University of Cambridge,
  Cambridge}, 26, 2009.

\bibitem{spall1998overview}
James~C Spall.
\newblock An overview of the simultaneous perturbation method for efficient
  optimization.
\newblock {\em Johns Hopkins apl technical digest}, 19(4):482--492, 1998.

\bibitem{Schuld_2019}
Maria Schuld, Ville Bergholm, Christian Gogolin, Josh Izaac, and Nathan
  Killoran.
\newblock Evaluating analytic gradients on quantum hardware.
\newblock {\em Physical Review A}, 99(3), mar 2019.

\bibitem{Banchi2021measuringanalytic}
Leonardo Banchi and Gavin~E. Crooks.
\newblock Measuring {A}nalytic {G}radients of {G}eneral {Q}uantum {E}volution
  with the {S}tochastic {P}arameter {S}hift {R}ule.
\newblock {\em {Quantum}}, 5:386, January 2021.

\bibitem{Wierichs2022generalparameter}
David Wierichs, Josh Izaac, Cody Wang, and Cedric Yen-Yu Lin.
\newblock General parameter-shift rules for quantum gradients.
\newblock {\em {Quantum}}, 6:677, March 2022.
\bibitem{PhysRevLett.119.180509}
Kristan Temme, Sergey Bravyi, and Jay~M. Gambetta.
\newblock Error mitigation for short-depth quantum circuits.
\newblock {\em {Phys. Rev. Lett.}}, 119:180509, November 2017.
\bibitem{https://doi.org/10.48550/arxiv.1012.1337}
Ran Cheng.
\newblock Quantum geometric tensor (fubini-study metric) in simple quantum
  system: A pedagogical introduction, 2010.

\bibitem{Stokes_2020}
James Stokes, Josh Izaac, Nathan Killoran, and Giuseppe Carleo.
\newblock Quantum natural gradient.
\newblock {\em Quantum}, 4:269, may 2020.

\bibitem{Gacon_2021}
Julien Gacon, Christa Zoufal, Giuseppe Carleo, and Stefan Woerner.
\newblock Simultaneous perturbation stochastic approximation of the quantum
  fisher information.
\newblock {\em Quantum}, 5:567, oct 2021.

\bibitem{PhysRevA.103.012405}
Andrea Mari, Thomas~R. Bromley, and Nathan Killoran.
\newblock Estimating the gradient and higher-order derivatives on quantum
  hardware.
\newblock {\em Phys. Rev. A}, 103:012405, Jan 2021.

\bibitem{haug2021optimal}
Tobias Haug and M.~S. Kim.
\newblock Optimal training of variational quantum algorithms without barren
  plateaus, 2021.
\bibitem{spall1997accelerated}
J.~C. Spall.
\newblock Accelerated second-order stochastic optimization using only function measurements.
\newblock In {\em Proceedings of the 36th IEEE Conference on Decision and Control},
  volume~2, pages 1417--1424, 1997.
\newblock doi: 10.1109/CDC.1997.657661.
\bibitem{github}
The code in this work can be accessed at the electronic address:
  https://github.com/nguyenvulinh666/Variational-Quantum-EigeinSolver.

\bibitem{RevModPhys.95.045005}
Zhenyu Cai, Ryan Babbush, Simon~C. Benjamin, Suguru Endo, William~J. Huggins, Ying Li, Jarrod~R. McClean, and Thomas~E. O'Brien.
\newblock Quantum error mitigation.
\newblock {\em {Rev. Mod. Phys.}}, 95:045005, December 2023.

\bibitem{mitarai2018quantum}
K.~Mitarai, M.~Negoro, M.~Kitagawa, and K.~Fujii.
\newblock Quantum circuit learning.
\newblock {\em Phys. Rev. A}, 98:032309, Sep 2018.

\end{thebibliography}
\end{document}